\documentclass[12pt]{article}

\usepackage[sort&compress,numbers]{natbib}
\usepackage[utf8]{inputenc}
\usepackage{amsmath,amssymb,mathtools}
\usepackage{graphicx}
\usepackage[hidelinks, pageanchor=false]{hyperref}
\usepackage{xcolor}
\usepackage{multirow}
\usepackage{babel}
\usepackage{xspace}
\usepackage{enumerate}
\usepackage{caption}
\usepackage{subcaption}
\usepackage{soul}
\usepackage{soul}
\usepackage{float} 
\usepackage{comment}
\usepackage{longtable}
\usepackage[
  a4paper,
  twoside=false,
  left=1.7cm,
  right=1.7cm,
  top=2.5cm,
  bottom=3cm
]{geometry}

\allowdisplaybreaks
 
\def\beq{\begin{equation}}
\def\eeq{\end{equation}}
\def\bea{\begin{align}}
\def\eea{\end{align}}

\def\Eq#1{Eq.~(\ref{#1})}

\def\id{\boldsymbol I}
\def\ra{\rangle}
\def\la{\langle}
\def\bra#1{\la #1|}
\def\ket#1{|#1\ra}
\newcommand{\braket}[2]{\la #1|#2\ra}

\begin{document} 
\begin{titlepage}
\renewcommand{\thefootnote}{\fnsymbol{footnote}}
\begin{flushright}
IFIC/22-14  \\  FTUV-22-0413.2034
\end{flushright}
\par \vspace{10mm}
\begin{center}
{\Large \bf
Quantum clustering and jet reconstruction at the LHC\\
}
\end{center}
\par \vspace{2mm}
\begin{center}
{\bf Jorge J. Mart\'{\i}nez de Lejarza, Leandro Cieri and Germ\'an Rodrigo}\footnote{Disclaimer: GR currently on leave at the European Research Council Executive Agency, European Commission, BE-1049 Brussels, Belgium. The views expressed are purely those of the writer and may not in any circumstances be regarded as stating an official position of the European Commission.},\\  
\vspace{5mm}

Instituto de F\'{\i}sica Corpuscular, Universitat de Val\`encia - Consejo Superior de Investigaciones Cient\'{\i}ficas, Parc Cient\'{\i}fic, E-46980 Paterna, Valencia, Spain


\vspace{5mm}

\end{center}

\par \vspace{2mm}
\begin{center} {\large \bf Abstract} \end{center}
\begin{quote}
\pretolerance 10000

Clustering is one of the most frequent problems in many domains, in particular, in particle physics where jet reconstruction is central in experimental analyses. Jet clustering at the CERN's Large Hadron Collider (LHC) is computationally expensive and the difficulty of this task will increase with the upcoming High-Luminosity LHC (HL-LHC).
In this paper, we study the case in which quantum computing algorithms might improve jet clustering by considering two novel quantum algorithms which may speed up the classical jet clustering algorithms. The first one is a quantum subroutine to compute a Minkowski-based distance between two data points, whereas the second one consists of a quantum circuit to track the maximum into a list of unsorted data. The latter algorithm could be of value beyond particle physics, for instance in statistics. When one or both of these algorithms are implemented into the classical versions of well-known clustering algorithms (\texttt{K-means}, Affinity Propagation and $k_T$-jet) we obtain efficiencies comparable to those of their classical counterparts. Even more, exponential speed-up could be achieved, in the first two algorithms, in data dimensionality and data length when the distance algorithm or the maximum searching algorithm are applied.

\end{quote}
\end{titlepage}
\setcounter{footnote}{0}
\renewcommand{\thefootnote}{\arabic{footnote}}

\section{Introduction}
\label{sec:intro}
Quantum computing devices, which are based on the laws of quantum mechanics, offer the possibi-lity to efficiently solve specific problems that become very complex or even unreachable for classical computers since they scale either exponentially or super-polynomially. Algorithms used in quantum computers~\cite{Feynman:1981tf} exploit the quantum principles of superposition and entanglement to clearly manifest a speed-up advantage over the classical counterpart algorithms. Two examples of these quantum algorithms are the well-known cases of Grover's database querying~\cite{Grover:1997fa} and Shor's factoring of integers into primes~\cite{Shor:1994jg}. These two quantum methods shown, for first time in the 1990s, a clear potential advantage over their corresponding classical analogues. In the past recent years, we have witnessed an impressively fast development of quantum computing algorithms going from optimization problems such as port-folios in fintech~\cite{Orus:2019xrh}, applications in quantum chemistry~\cite{Liu:2020eoa}, nuclear physics and Monte Carlo simulation~\cite{Holland:2019zju,Lynn:2019rdt,montanaro:2015}, combinatorial optimization~\cite{Kokail:2018eiw}, to state diago\-nalization~\cite{Larose:2019,PhysRevA.101.062310}.

Very recently, quantum algorithms have started to be applied in solving problems which appear in high-energy particle physics~\footnote{For a recent review on the applications of quantum computing to data analysis in HEP we refer the reader to Ref.~\cite{Delgado:2022tpc} and references therein.} (HEP). The data already taken at the CERN’s Large Hadron Collider (LHC) and its upcoming Run 3 (which is scheduled to start in the spring of 2022) demand intense data analysis routines and very precise theoretical predictions~\cite{EPPPG:2019qin} which are computationally very expensive. This situation will be even more challenging in the posterior high-luminosity phase of the LHC (HL-LHC)~\cite{Gianotti:2002xx} and the planned future colliders~\cite{FCC:2018byv,Roloff:2018dqu,CEPCStudyGroup:2018ghi}. Recent applications of quantum algorithms to HEP cover diverse subareas such as jet clustering ~\cite{Wei:2019rqy,Pires:2021fka,Pires:2020urc}, jet quenching~\cite{Barata:2021yri}, determination of parton densities~\cite{Perez-Salinas:2020nem}, simulation of parton showers~\cite{Bauer:2019qxa,Williams:2021lvr,Bepari:2020xqi}, heavy-ion collisions~\cite{DeJong:2020riy}, quantum machine learning~\cite{Guan:2020bdl,Wu:2020cye,Felser:2020mka,Abel:2022lqr,Araz:2022haf,Ngairangbam:2021yma,Araz:2021zwu,Blance:2020nhl}, lattice gauge theories~\cite{Jordan:2012xnu,Banuls:2019bmf,Zohar:2015hwa,Byrnes:2005qx} and multi-loop Feynman integrals~\cite{Ramirez-Uribe:2021ubp,Ramirez-Uribe:2021drs}.

In the present paper we address the problem of clustering and jet reconstruction from collision data, which is a nontrivial and computationally expensive task, as it often involves performing optimizations over potentially large numbers of final-state particles. To give a rough idea of how demanding this activity is, the state-of-the-art algorithm in jet clustering needs few months to clusterize all the particles generated in the data of interest that is produced at the LHC in just one year \cite{Evans2009TheLH}. Moreover, with the upcoming HL-LHC, the number of events will be up to an order of magnitude more than in earlier runs \cite{HEPSoftwareFoundation:2017ggl} and also the pile-up (simultaneous proton-proton collisions per bunch crossing) will increase by a factor of 5 \cite{Collaboration:2802918}. Therefore, the state-of-the-art algorithm will require roughly 50 times the computational time needed now. So we would be talking about a few tens of years for processing the data of interest generated in just a year. This evidences the necessity of developing fast and effective jet clustering algorithms. 

With this in mind, we consider the possibility of using quantum algorithms to improve the velocity in jet identification. Here we focus on three well-known classical algorithms: the \texttt{K-means} clustering~\cite{macqueen1967some,ball1967clustering}, the Affinity Propagation (\texttt{AP}) algorithm~\cite{Frey2007ClusteringBP} and the $k_T$-jet clustering method in all its variants~\cite{Catani:1991hj,Catani:1993hr,Ellis:1993tq,Cacciari:2008gp,Dokshitzer:1997in}. We propose the correspon-ding quantum versions of the precedents algorithms: quantum \texttt{K-means} clustering, quantum \texttt{AP}-algorithm and quantum $k_T$-based algorithms.

Clustering is one of the most frequent classic problems in machine learning and computational geometry. It is a major data analysis tool used in such domains as marketing research, data mining, bioinformatics, image processing, pattern recognition and also in HEP. The popular \texttt{K-means} formulation~\cite{ball1967clustering,macqueen1967some}, which is a method of vector quantization originally proposed for signal processing, involves the partition of $n$ observations into $K$ clusters in which each observation belongs to the cluster with the nearest mean (cluster center or cluster centroid), serving as a prototype of the cluster. Solving this problem exactly is NP-hard~\footnote{NP-hard problems are not solvable in polynomial time but can be verified in polynomial time.} (Non-deterministic Polynomial-time hardness), even with just two clusters~\cite{Drineas:2004fa}. Forty years ago, Lloyd~\cite{Lloyd:1982ls} proposed a local search solution that is still very widely used today. Usually referred to simply as \texttt{K-means}, Lloyd’s algorithm begins with $K$ arbitrary centers, typically chosen uniformly at random from the data points. Each point is then assigned to the nearest center, and each center is recomputed as the center of all points assigned to it. These two steps (assignment and center calculation) are repeated until the process stabilizes.

The improved version of the \texttt{K-means} method, the \texttt{K-means++} algorithm~\cite{K-Means++}, initializes the \texttt{K-means} algorithm by choosing random starting centers with very specific probabilities. This strategy outperforms \texttt{K-means} in terms of both accuracy and speed, often by a substantial margin \cite{K-Means++}. \texttt{K-means} is a method of cluster analysis using a pre-specified number of clusters. It requires an advance (\textit{a priori}) knowledge of $K$ and belongs to the group of the so-called \textit{partitional clustering algorithms}. The classical \texttt{K-means} algorithm has already been used in high-energy physics in Refs.~\cite{Chekanov:2005cq,Thaler:2011gf,Stewart:2015waa,Wong:2018frb}.
For example, in Ref. \cite{Chekanov:2005cq}, the use of \texttt{K-means} led to 25 \% and 40\% improvement of the top quark and $W$ boson mass resolution, respectively, compared to the $k_T$ (Durham) algorithm, and reduced the systematic uncertainty in the measured peak positions. 
As a drawback, \texttt{K-means} was roughly three times slower than the Durham algorithm,
therefore the interest to explore potential speed ups. In Ref.~\cite{Thaler:2011gf}, the tagging performance of $N$-subjetiness for boosted top quarks was improved through minimization using a variant of \texttt{K-means}. The XCone jet algorithm introduced in Ref.~\cite{Stewart:2015waa} is closely related to the traditional \texttt{K-means} and its variants. Finally, \texttt{K-means} has been used in Ref.~\cite{Wong:2018frb} to identify minijets at low $p_T$.

The \texttt{AP} algorithm, is a clustering method that identifies representative examples (exemplars) within a given dataset by exchanging messages between all data points. Points are then grouped with their most representative exemplar to give the final set of clusters. The \texttt{AP} algorithm has been successfully applied to a wide range of problems including face recognition, gene identification, putative exons using microarray data 
\cite{Leone_2007, Sumedha_2008,Bailly_Bechet_2009} and astrophysics \cite{GonzlezMartn2017}.
In high-energy physics, it has been used to cluster replicas of parton densities \cite{Carrazza:2016sgh}. In Ref. \cite{Frey2007ClusteringBP}, it was shown that \texttt{AP} might be faster and more accurate than the \texttt{K-means} \cite{ball1967clustering,macqueen1967some} clustering algorithm in solving certain problems. The \texttt{AP} algorithm is solid and well understood and the number of clusters is not needed to be pre-specified. Among its disadvantages, the high time complexity turns out to make it not suitable for very large datasets, and the clustering result is typically sensitive to the parameters involved in the \texttt{AP} algorithm.
Our motivation in using it for jet clustering comes from the fact that it does not need the number of clusters to be defined beforehand.

Hierarchical clustering also known as hierarchical cluster analysis (HCA) is also a method of cluster analysis that seeks to build a hierarchy of clusters without having an \textit{a priori} fixed number of clusters. The $k_T$-based algorithms~\cite{Cacciari:2011ma} belong to the hierarchical category, which needs a linkage function that defines the distance between any two sub-sets (and relies on the base distance between elements). It is the most widely used jet clustering algorithm in the LHC experiments.

The quantum \texttt{K-means} clustering algorithm was presented in Refs.~\cite{Blance:2020ktp,Pires:2021fka} for HEP. An earlier study of the quantum \texttt{K-means} can be found in Ref.~\cite{Abhi:2020}. Both implementations make use of the Euclidean distance to perform the clustering of particles. In this paper, we present a version of the quantum \texttt{K-means} clustering algorithm which is based on the definition of a Minkowskian distance at the quantum level for the first time. Considering the case of the quantum version of the \texttt{AP} algorithm, it uses the invariant sum squared as a metric in the similarity matrix and calculates it through a quantum subroutine with a similar procedure as in the quantum \texttt{K-means} implementation. Regarding the quantum $k_T$-based algorithms, to our knowledge, it is the first time it has been presented in the literature. In addition, the search for the maximum distance used in our implementation is performed with a new quantum algorithm. This new quantum method is presented in a general way, and we comment on its reach regarding other areas of interest. Beyond the specific application to jet clustering, the quantum algorithms presented in this paper are of interest to the particle physics and quantum computing communities.

This paper is organized as follows. In Section \ref{sec:Qdistances} we introduce our notation and we define the Euclidean and Minkowskian quantum distances. In Section \ref{sec:qsearching} we present our new quantum algorithm in order to search the maximum in a set of a given number of elements. We consider the quantum version of the \texttt{K-means} clustering, \texttt{AP} and $k_T$-based algorithms in Section \ref{sec:jetalgorithms}. In Section \ref{sec:Qsimul} we present our results considering the quantum simulations of these algorithms and a proof-of-concept implementation with Gaussian datasets as well as with simulated LHC physical events. We also compare their performance in detail. We discuss their differences and conceptual similarities and we compare them with their classical counterparts. A brief summary of our results is presented in Section \ref{sec:conclu}.

\section{Quantum distances}
\label{sec:Qdistances}

In quantum computing, it is essential to have the ability to measure quantum entanglement between two states, as in many cases it determines the possibility of obtaining a quantum advantage~\cite{Foulds:2021}.  We rely on the \textit{SwapTest} method~\cite{Buhrman:2001} (see Appendix~\ref{app:swaptest} for more details) in order to probe the entanglement between two given states. The definition of quantum distances (Euclidean distance or Minkowski invariant sum squared) presented in this Section, makes use of the \textit{SwapTest} procedure.
 
\subsection{Euclidean quantum distance}

We start by considering $N$ data points or vectors
in an Euclidean $d$-dimensional space, 
$\{{\bf x}_i\}_{i=1,\ldots, N}$, which
are encoded as quantum states of the form 
\beq
\ket{x_i} = |{\bf x}_i|^{-1} \sum_{\mu=1}^{d} x_{i,\mu} \, \ket{\mu}~,
\eeq
where $|{\bf x}_i| = \sqrt{\sum_{\mu=1}^d (x_{i,\mu})^2}$ is the modulus 
of the vector ${\bf x}_i$, and $x_{i,\mu}$ are its components.  
Each vector requires $n \ge \log_2 d$ qubits to be encoded, i.e. for $d=3$ we need two entangled qubits where one of its states remains free and is not used. The Euclidean distance between two vectors ${\bf x}_i$ and ${\bf x}_j$ is defined classically as
\beq
\label{eq:EuclidDist}
d_E^{\rm (C)}({\bf x}_i,{\bf x}_j) = |{\bf x}_i - {\bf x}_j|~,
\eeq
where the subscript $E$ stands for Euclidean and the superscript ${\rm C}$ denotes that it corresponds to the classical version.

The quantum analogue of Eq.~\eqref{eq:EuclidDist} is obtained by using the controlled \textit{SwapTest} method. In order to define the Euclidean quantum distance between the $d$-dimensional vectors ${\bf x}_i$ and ${\bf x}_j$, we entangle the corresponding associated quantum states $\ket{x_i}$ and $\ket{x_j}$, and define the following subsidiary states
\beq
\ket{\psi_1} =\frac{1}{\sqrt{2}} \left( \ket{0, x_i} + \ket{1, x_j} \right)~, \qquad
\ket{\psi_2} =\frac{1}{\sqrt{Z_{ij}}} \left( |{\bf x}_i| \ket{0} -|{\bf x}_j| \ket{1} \right)~,
\label{eq:varphi}
\eeq
where $Z_{ij}=|{\bf x}_i|^2+|{\bf x}_j|^2$ 
is a normalization factor and $\ket{0}$ and $\ket{1}$ are the states of an ancillary qubit. It is also convenient to define the 
swapped state $\ket{\psi'_1}$
\beq
\ket{\psi_1'} =\frac{1}{\sqrt{2}} \left( \ket{x_i, 0} + \ket{x_j, 1}  \right)~.
\label{eq:psiprime}
\eeq
The inner products between the quantum states defined in Eqs. \eqref{eq:varphi} and \eqref{eq:psiprime} are written as follows
\beq
\braket{\psi_1'}{\psi_2} = \frac{1}{\sqrt{2Z_{ij}}} 
\left(|{\bf x}_i| \bra{x_i} - |{\bf x}_j| \bra{x_j} \right)~, \qquad
\braket{\psi_2}{\psi_1} =\frac{1}{\sqrt{2Z_{ij}}} 
\left(|{\bf x}_i| \ket{x_i} - |{\bf x}_j| \ket{x_j} \right)~.
\label{eq:Edistance}
\eeq
From where
\beq
\braket{\psi_1'}{\psi_2} \braket{\psi_2}{\psi_1} 
= \frac{1}{2Z_{ij}}|{\bf x}_i-{\bf x}_j|^2~.
\label{eq:overlap}
\eeq
Therefore (see \Eq{eq:p0swaptest} in Appendix~\ref{app:swaptest}\-), the 
Euclidean quantum distance is
\beq
d_E^{\rm (Q)}({\bf x}_i, {\bf x}_j) = \sqrt{2Z_{ij}(2P_{\Psi_3}(\ket{0})-1)}\,,
\label{eq:qdistance}
\eeq
where the superscript $Q$ refers to the \textit{Quantum} version of the distance $d_E$ and the subscript $\Psi_3$ in the probability $P$, means that it is considered the resulting probability of measuring the ancillary qubit in the state $\ket{0}$ in the last of the three steps in the \textit{SwapTest} procedure.

\subsection{Quantum invariant sum squared in Minkowski space}
\label{subsec:qinvmass}

Vectors in high-energy physics are defined in a
four-dimensional space-time with Minkowski metric. They have 
the form $x_i = (x_{i,0}, {\bf x}_i)$, where $x_{i,0}$ is the 
temporal component and ${\bf x}_i$ represent the three spatial components. In the following, we assume that the dimension of the space-time is $d$, where $d-1$ is the number of spatial components. We shall define the analogue of the Euclidean classical distance in the Minkowski space corresponding to the invariant sum squared $s_{ij}^{\rm (C)}$, which is commonly called invariant mass squared when vectors are particle four-momenta,
\beq
s_{ij}^{\rm (C)} = (x_{0,i}+x_{0,j})^2 
- |{\bf x}_i + {\bf x}_j|^2~. 
\eeq

This quantity, which is Lorentz invariant, can be used as test distance to measure similarity between particle momenta.  It is also equivalent to the distance used in some of the traditional jet-clustering algorithms at $e^+e^-$ colliders~\cite{JADE:1982ttq,Bethke:1991wk,Rodrigo:1999qg}. It is necessary to apply twice the \textit{SwapTest} subroutine (presented in Appendix A) for computing the Minkowski-type distance through a quantum algorithm. Once for the spatial and once for the temporal components.

The spatial distance is computed through the procedure explained in the previous section with a slight modification with respect to \Eq{eq:Edistance} (change of sign in the term proportional to qubit $\ket{1}$)
\beq
\ket{\psi_2} \longrightarrow \ket{\psi_2}  =\frac{1}{\sqrt{Z_{ij}}} \left( |{\bf x}_i| \ket{0} +|{\bf x}_j| \ket{1} \right)~,
\eeq
whereas the temporal distance is computed as a result of the overlap of the following states:
\begin{equation}
\ket{\varphi_1} = H \ket{0} = \frac{1}{\sqrt{2}} \left( \ket{0} + \ket{1} \right)~, \qquad 
\ket{\varphi_2} =\frac{1}{\sqrt{Z_0}} \left(x_{0,i} \ket{0} +x_{0,j} \ket{1} \right)~,
\label{eq:minkstates}
\end{equation}
where $Z_{0}=x_{0,i}^2+x_{0,j}^2$. Then, applying the \textit{SwapTest} to these states one gets the relation:
\begin{equation}
P(|0\rangle|_{time})=\frac{1}{2}+\frac{1}{2} |\langle \varphi_1| \varphi_2 \rangle|^2 \ ,
\label{eq:p0time}
\end{equation}
where the overlap $|\langle \varphi_1| \varphi_2 \rangle|^2$ is trivially given by
\begin{equation}
|\langle \varphi_1| \varphi_2 \rangle|^2 = \frac{1}{2Z_0}(x_{0,i} +x_{0,j})^2.
\label{eq:hphioverlap}
\end{equation}
Therefore:
\begin{equation}
(x_{0,i}+x_{0,j})^2 =2 Z_{0}(2P_{\Psi_3}(|0\rangle|_{time})-1)~.
\label{eq:parttime}
\end{equation}
At this point, the quantum version of the invariant sum squared follows from the combination of results from \Eq{eq:qdistance} and \Eq{eq:parttime}:
\beq
s_{ij}^{\rm (Q)} =2\big( 
Z_0(2P_{\Psi_3}(|0\rangle|_{time})-1)-Z_{ij}(2P_{\Psi_3}(|0\rangle|_{spatial})-1)\big).
\label{eq:qdistancemink}
\eeq

\begin{figure}[th!]
    \centering
    \includegraphics[width=0.6\textwidth]{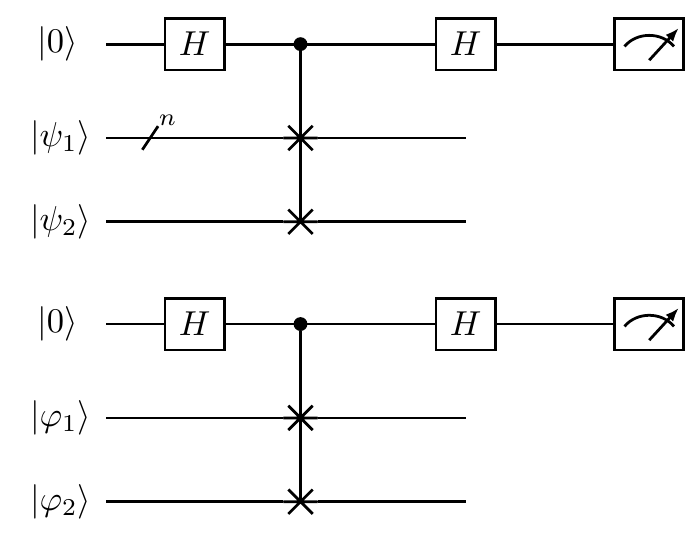}
    \caption{Quantum circuit to obtain the invariant sum squared between two $d$-dimensional vectors in Minkowski space.}  
    \label{fig:qsmassinvariant}
\end{figure}

The quantum circuit used to implement the invariant sum-squared distance is shown in Fig.~\ref{fig:qsmassinvariant}. 

In the first three wires, the \textit{SwapTest} is applied to the spatial  components, where we assume that the states $\psi_1$, $\psi_2$ have been loaded from a quantum Random Access Memory (qRAM) in $\mathcal{O}(\log (d-1))$, since the state $\psi_1$ is encoded in $\log_2 (d-1)$ qubits. On the other hand, from the fourth wire onward, the \textit{SwapTest} is applied to the temporal components. In this case, it takes $\mathcal{O}(1)$, since we only have 1-dimensional qubit states.

\section{Quantum maximum search by amplitude encoding}
\label{sec:qsearching}

Finding a particular member belonging to a dataset is a recurring problem in data analysis. This is a computationally very expensive task. However, quantum computing offers suitable tools to solve data query in a shorter computational time. In particular,  it is well known the quadratic speed up exhibited by Grover's algorithm~\cite{Grover:1997fa}.
In this paper, we present a considerably simpler algorithm that is used exclusively to find the maximum in a list of values. This algorithm, although very elementary, is sufficiently accurate for the applications that we will present in Sections~\ref{subsec:qkmeans} and~\ref{subsec:qktalgorithm}. To our knowledge, it is the first time presented in the literature.

\begin{figure}[th!]
    \centering
    \includegraphics[width=0.9\textwidth]{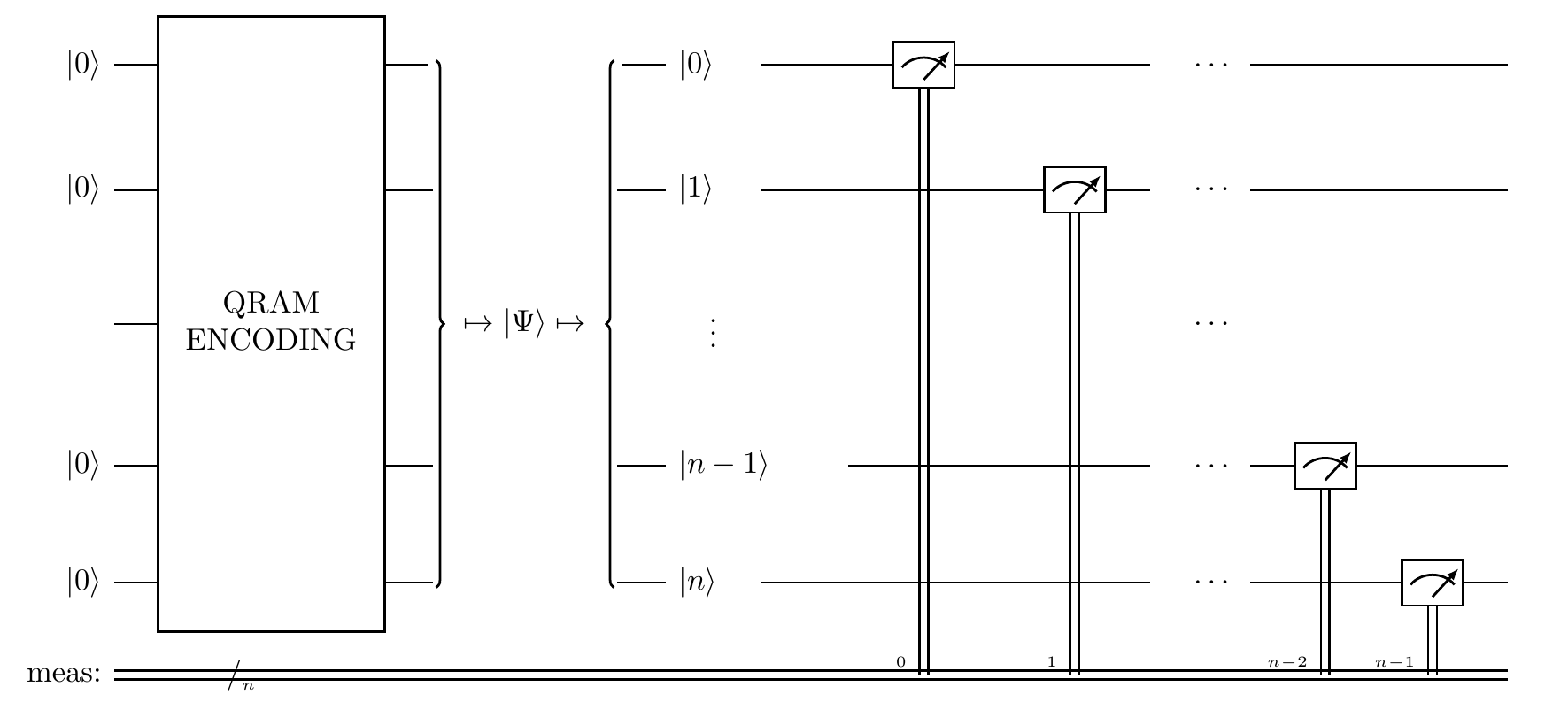}
    \caption{Quantum circuit for quantum maximum searching by qRAM amplitude encoding.}  
    \label{fig:qsearch}
\end{figure}

Let $L[0, \ldots ,N-1]$ be an unsorted list of $N$ items. Solving the maximum searching problem is to find the index $y$ such that $L\left[y\right]$ is the maximum. The quantum algorithm to solve that problem using amplitude encoding proceeds in two steps:
\begin{enumerate}
    \item  The list of $N$ elements is encoded into a $\log_2(N)$ qubits state as follows:
    \begin{equation}
    \ket{\Psi} = \frac{1}{\sqrt{L_{sum}}}\sum_{j=0}^{N-1}L\left[j\right] \, \ket{j}~,
        \label{eq:amplitude encoding}
    \end{equation}
    where $L_{sum}=\sum_{j=0}^{N-1}L[j]^2 $ is a normalization constant.
    This amplitude encoding is achieved using qRAM.
    \item  The final state is measured. This step is rerun several times to reduce the statistical uncertainty. Once done, the most repeated state gives us the maximum.
\end{enumerate}
The graphical representation of the algorithm is shown in Fig.~\ref{fig:qsearch}, where $n=\log_2(N)$ qubits are needed to encode a list of $N$ (real) elements. 

The bottleneck of this procedure underlies in encoding data into a quantum state. Assuming data is stored in a qRAM, as would be the case on a true universal quantum computer, encoding takes $\mathcal{O}(\log_2(N))$ steps~\cite{lloyd2013quantum,2008QRAM,2008Arch,demartini2009experimental,2010Mag,PhysRevLett.105.140501,PhysRevLett.105.140502,PhysRevLett.105.140503}. The corresponding classical algorithms typically used to obtain the minimum of an unsorted list of $N$ items are of order $\mathcal{O}(N)$. Therefore, with the assumptions considered,  the improvement introduced by this quantum algorithm is exponential.

The well-known quantum minimum searching algorithm proposed by Dürr and Høyer \cite{Durr:1996nx} is $\mathcal{O}(\sqrt{N})$. After their theoretical paper \cite{Durr:1996nx} the algorithm was studied and implemented in a quantum simulator (see Ref.~\cite{DBLP:journals/corr/abs-1804-03719}). In summary, previous implementations~\cite{DBLP:journals/corr/abs-1804-03719} of the Dürr and Høyer algorithm suggests that it could be improved, given the excessive number of qubits needed to implement the method, the unviability to hard code a different oracle for each element, the large number of \textit{shots} required and (in some cases) the poor performance obtained. This is the aim of the new quantum maximum searching algorithm by amplitude encoding through qRAM presented here: the improvement of the previous enumerated challenges.

Nevertheless, the new algorithm presented in this paper and the corresponding Dürr and Høyer quantum method share common features that could lead to miss-identification of the respective absolute maximum and minimum. These cases, in which the list typically presents a very low standard deviation (or the largest/minimum values are very close to each other) could manifest difficulties related to the fact that the probability of measuring several candidates would be almost identical.

Regarding the practical implementation of the quantum algorithm presented in this paper, the results shown in Section \ref{sec:Qsimul} reveal that these potential difficulties do not manifest strongly in the context of jet clustering.

Beyond the jet clustering procedure in HEP, there are other fields where our quantum algorithm could be of value. For instance, in the so-called Extreme Value Theory (EVT)~\cite{smith1990extreme}.
According to Gumbell 1958~\cite{Gumbel1958}, this particular field studies the probability distribution of the desired data by focusing on the outliers with the ultimate goal of being able to predict them in the future. It is precisely in this estimation of the extreme values where our algorithm could be useful. Since for the predictive models historical data has to be analysed and therefore extreme values have to be searched in large data lists. This would mean that our algorithm could be implemented successfully in statistical analysis of extreme data, including actuarial and financial sciences, meteorology, material sciences, engineering and environmental sciences climatology, geology, hydrology and highway traffic analysis~\cite{Reiss2007,Coles2001,castillo2004}.

\section{Quantum clustering algorithms}
\label{sec:jetalgorithms}

\subsection{\texttt{K-means} algorithm}
\label{subsec:kmeans}

\texttt{K-means} is an unsupervised machine learning algorithm that classifies the elements of a dataset into $K$ groups called clusters~\cite{ball1967clustering,macqueen1967some}. The data points within each cluster have to be as similar (near) as possible whereas the clusters themselves have to be as different (far) as possible from each other. 
The input for this algorithm is a set of $N$ data points or vectors, in $d$ dimensions as well as the number of clusters $K$, with $K\le N$, and its output is a set of $K$ centroids, calculated by averaging the position of the data points corresponding to each group, thus defining $K$ clusters The flow chart of this algorithm is the following:
\begin{enumerate}
    \item $K$ initial centroids within the data points are generated. They can be generated randomly or through a specific method such as \textit{kmeans}$++$~\cite{K-Means++}.
    \item Each data point is assigned to its closest centroid according to a distance that has been defined in advance, thus the $K$ clusters are defined. The most commonly used distance is the Euclidean distance.
    \item Each centroid is recalculated by averaging the associated data points.
    \item Steps 2 and 3 are repeated until all centroids stabilize and convergence is achieved.
\end{enumerate}

This \texttt{K-means} algorithm has a sophisticated quantum version that differs from its classical counterpart in two points~\cite{kopczyk2018quantum}. First, the quantum \texttt{K-means} introduces a quantum method to calculate the distance between data points. Second, the quantum version also includes a procedure for obtaining the minimum distance of each data point with respect to the $K$ centroids, which is achieved by Dürr and Høyer's algorithm~\cite{Durr:1996nx}.

In this paper, we focus on a new quantum version of the \texttt{K-means} algorithm, where the calculation of distances is made quantumly and the minimum distance of each data point to the centroids is obtained with the quantum maximum searching algorithm\footnote{We may apply this algorithm for finding the minimum since obtaining the minimum amongst the distances is equivalent to obtaining the maximum of their inverses: $s_{ij}^{-1}$.} explained in Section~\ref{sec:qsearching}. 
Other quantum versions of the \texttt{K-means} algorithm have been studied in Refs.~\cite{Blance:2020ktp,Pires:2021fka} and~\cite{Abhi:2020}, where an Euclidean distance was used to separate the particles from each other. In this paper, we analyse for the first time an implementation of the \texttt{K-means} algorithm with a Minkowski-type quantum distance, as defined in Section~\ref{subsec:qinvmass}.

The time complexity of this algorithm is estimated by analysing the time complexity of its components. The distances that have to be calculated are $\mathcal{O}(N)$, the search of a minimum distance for every data point with respect to the centroids would be $\mathcal{O}(\log K)$, and the calculation of each distance itself would require $\mathcal{O}(\log (d-1))$ qubits assuming the data is stored in a qRAM. This results in a speedup from $\mathcal{O}(NKd)$ in the classical version to $\mathcal{O}(N\log K \log(d-1))$ in our quantum version. Therefore an exponential speed-up in the  number of clusters and in the vector dimensionality would be achieved. A quantum simulation of the quantum \texttt{K-means} algorithm is presented in Section~\ref{subsec:qkmeans}.

\subsection{Affinity Propagation algorithm}

Although \texttt{K-means} is a successful algorithm capable of clustering data in a satisfactory manner, it needs the number of clusters $K$ to be defined beforehand, which is not typically the case in HEP applications. The Affinity Propagation (\texttt{AP}) algorithm \cite{Frey2007ClusteringBP}, which is an unsupervised machine learning algorithm, does not need the number of clusters as an input. \texttt{AP} only takes as input the data points that have to be classified. So,
let $x_1, \ldots ,x_N$ be a set of data points.
Then, a function $s$ to quantify the similarity between points is computed. In such a way that $s(i,j)\geq s(i,k)$ if and only if $x_i$ is more similar to $x_j$ than to $x_k$. The most common metrics to measure the similarity is the negative squared distance of the two points we are comparing: $s(i,j)=-|x_i-x_j|$. 
The diagonal $s(i,i)$ of the matrix $s$ is especially relevant since it stores values referred as ``preferences" that are related to how likely a particular instance is to become an exemplar, i.e, a cluster. Most of the metrics make the diagonal $s(i,i)$ be $s(i,i)=~0, \, \forall i\leq N$, although it can be different from $0$. Hence, on the first iteration, every element $s(i,i)$ is set to the same certain value, which is typically the median similarity of all pairs of inputs. 
Next, two matrices are calculated that are related to the concept of message exchanging between data points \cite{Frey2007ClusteringBP}.
First, there is the responsibility matrix $R$. This matrix contains the values $r(i,k)$ that quantify the suitability of point $k$ to serve as the exemplar for point $i$, compared to other candidate exemplars for $i$. Then comes the availability matrix $A$, whose elements $a(i,k)$ reflect how appropriate it would be for point $i$ to select point $k$ as its exemplar, relative to the preferences of other points for $k$ as an exemplar. As they have been described, both matrices could be viewed as log-probability ratios. Then, the \texttt{AP} flow chart reads:
\begin{enumerate}
    \item The matrices $R$ and $A$ are initialized to zero.
    \item The responsibility matrix is computed:
    \beq
    r(i,k)= s(i,k)-\max_{q\neq k}\lbrace a(i,q)+s(i,q)\rbrace.
    \eeq
    
    \item The availability matrix is computed:
    \beq
    a(i,k)= \min \left( 0, r(k,k)+\sum_{q\notin \lbrace i,k \rbrace}\max(0,r(q,i))\right) \mathrm{for} \, i \neq k, \, \mathrm{and}
    \eeq
    \beq
    a(k,k)= \sum_{q\neq k} \max(0,r(q,k).
    \eeq
    \item Steps 2 and 3 are repeated until either the cluster boundaries remain unchanged for several iterations, or a predetermined number (of iterations) is reached.
\end{enumerate}

Once convergence has been reached, the exemplars i.e, the clusters, are obtained from the final matrices as those whose $r(i,i)+a(i,i)>0$.
This algorithm takes $\mathcal{O}(N^2)$ steps to fill the similarity matrix, and also computing each element takes $\mathcal{O}(d)$, since a distance between two $d$-dimensional points has to be calculated. Moreover, steps 2 and 3 are repeated a number $T$ of times, so the final time complexity of this algorithm is $\mathcal{O}(N^2Td)$.

Here, a quantum (hybrid) algorithm is presented which uses the invariant sum squared as a metric in the similarity matrix and calculates it through a quantum subroutine, as the \texttt{K-means} algorithm described in the subsection \ref{subsec:kmeans}. Then, a speedup would be achieved, since computing the distances only requires $\mathcal{O}(\log(d-1))$ qubits. So, the quantum \texttt{AP} algorithm, which is as far as we know completely original, would have a time complexity of $\mathcal{O}(N^2T\log(d-1))$.

\subsection{Generalised $k_T$-jet algorithm}
\label{subsec:ktjetalgorithm}

The inclusive variant of the generalised $k_T$-jet algorithm is formulated as follows \cite{Cacciari:2011ma}:
\begin{enumerate}
    \item For each pair of partons $i$, $j$ the following distance is computed:
    \begin{equation}
        d_{ij}=\mathrm{min}  (p_{T,i}^{2p},p_{T,j}^{2p}) \Delta R_{ij}^2/R^2,
        \label{eq:kt}
    \end{equation}
    with $\Delta R_{ij}^2=(y_i-y_j)^2+(\phi_i-\phi_j)^2$, where $p_{T,i}$, $y_i$ and $\phi_i$ are the transverse momentum (with respect to the beam direction), rapidity and azimuth of particle $i$. $R$ is a jet-radius parameter usually taken of order 1. For each particle $i$ the beam distance is $d_{iB}=p_{T,i}^{2p}$.
    \item Find the minimum $d_{min}$ amongst all the distances $d_{ij}$, $d_{iB}$. If $d_{min}$ is a $d_{ij}$, the particles $i$ and $j$ are merged into a single particle summing their four-momenta (this is the E-scheme recombination); if $d_{min}$ is a $d_{iB}$ then the particle $i$ is declared as a final jet and it is removed from the list.
    \item Repeat from step 1 until there are no particles left.
\end{enumerate}

It is noticeable that for specific values of $p$ in Eq.~\eqref{eq:kt}, the generalised $k_T$ algorithm is reduced to the algorithms: $k_T$ ($p=1$), Cambridge/Aachen ($p=0$) and anti-$k_T$ ($p=-1$).
As it is claimed in Ref. \cite{2006}, this classical version of the $k_T$-jet algorithm is $\mathcal{O}(N^3)$, since the bottleneck of the algorithm is scanning the $\mathcal{O}(N^2)$ table with all the distances $d_{ij}$, $d_{iB}$, and it has to be done $N$ times. 
Nevertheless, the \texttt{FastJet} algorithm is able to reduce the complexity to ${\cal O}(N^2)$. It is achieved by identifying each particle's geometrical nearest neighbour, thereby it is not necessary to construct a size-$N^2$ table of $d_{ij}$, but only the size-$N$ array, $d_{i\mathcal{G}_i}$, where $\mathcal{G}_i$ is $i$'s geometrical nearest neighbour. 
Furthermore, this \texttt{FastJet} algorithm can be optimized further using the so-called Voronoi diagrams achieving a reduction in the time complexity from
$\mathcal{O}(N^2)$ to $\mathcal{O}(N \log N)$.

Regarding the quantum version of this algorithm, the distance $\Delta R_{ij}^2$ will be computed classically whereas the minimum will be obtained through a quantum algorithm. This is due to the fact that the speed up achieved by obtaining the minimum here with a quantum subroutine will be dominant. Thereby, what is to be used here is the new algorithm to obtain the maximum of a list of values (see Section~\ref{sec:qsearching}). So obtaining the minimum amongst all the distances $d_{ij}$, $d_{iB}$ will turn out to be obtaining the maximum of its inverses: $d_{ij}^{-1}$, $d_{iB}^{-1}$. Actually, these inverse distances are what will be computed directly for each pair $i$, $j$. Since computing the distances and thereafter computing its inverses would require traversing a vector of size $N$, so it would have a complexity $\mathcal{O}(N^2)$ . With that in mind one may also directly compute $d_{ij}^{-a}$, $d_{iB}^{-a}$, with $a \in \mathbb N$, to increase the separation among the data, which makes the maximum more likely when measuring. And this will not increase the overall time complexity of the algorithm either. In Section~\ref{sec:Qsimul} we compare the results obtained when applying the algorithm with different $a$ values.

The quantum maximum searching algorithm presented above could be applied to the $k_T$-jet algorithm successfully because  accuracy is not critical. Even if our quantum algorithm fails to obtain the absolute maximum in one of the multiples iterations, this could end up not affecting the overall jet clustering process. Since an error in finding the maximum will provoke a flip in the order in which two particles merge, and the final result will in many cases be independent of this permutation. 

As a final remark, notice that the $k_T$-jet quantum algorithm would be $\mathcal{O}(N^2\log(N))$, since computing all the distances takes $\mathcal{O}(N^2)$ and finding the minimum would be $\mathcal{O}(\log(N))$, in comparison with the $\mathcal{O}(N^3)$ that requires its classical analogue \cite{2006}.
Furthermore, the quantum minimum searching could also be implemented in the \texttt{FastJet} algorithm of complexity ${\cal O}(N^2)$. In this case, the resulting quantum algorithm would be $\mathcal{O}(N\log(N))$, which is of the same order as the \texttt{FastJet} algorithm version with  Voronoi diagrams, which is the most efficient clustering algorithm known to date. This quantum \texttt{FastJet} algorithm has been tested in Section \ref{subsec:ktjetalgorithm} with LHC physical datasets.

\section{Quantum simulations}
\label{sec:Qsimul}

The implementation of the quantum algorithms has been performed through the open-source IBMQ software. In particular, the Python module \textit{Qiskit} developed by IBMQ has been used to build the quantum circuit to calculate the invariant sum squared as described in Section \ref{subsec:qinvmass} for the \texttt{K-means} and the \texttt{AP} algorithm, as well as to build the quantum circuit for finding the minimum distance in the \texttt{K-means} and the $k_T$-jet algorithm. Afterward, these quantum subroutines have been introduced into their respective classical algorithm substituting the classical part they are speeding up. The \textit{Qiskit} module serves for executing circuits on real quantum devices. Nevertheless, in previous studies such as~\cite{Abhi:2020} and~\cite{Fanizza:2019}, it has been found that the experimental error associated with the quantum devices provided by IBMQ is not yet sufficiently small to extract significant results. Hence, the  algorithms presented here have been executed on a quantum simulator that offers an unrestricted and noise-free environment.
A quantum implementation in an existing quantum device taking advantage of the claimed maximum speed-up is also not possible, as a qRAM architecture does not exist yet. Nonetheless, the quantum simulations in IBMQ presented in this section show a satisfactory performance and clustering efficiencies comparable to those of their classical counterparts.

\subsection{Quantum \texttt{K-means} with Minkowski-type distance}
\label{subsec:qkmeans}

\begin{figure}[ht!]
       \centering
\begin{subfigure}{.49\textwidth}
  \centering
  \includegraphics[width=.99\linewidth]{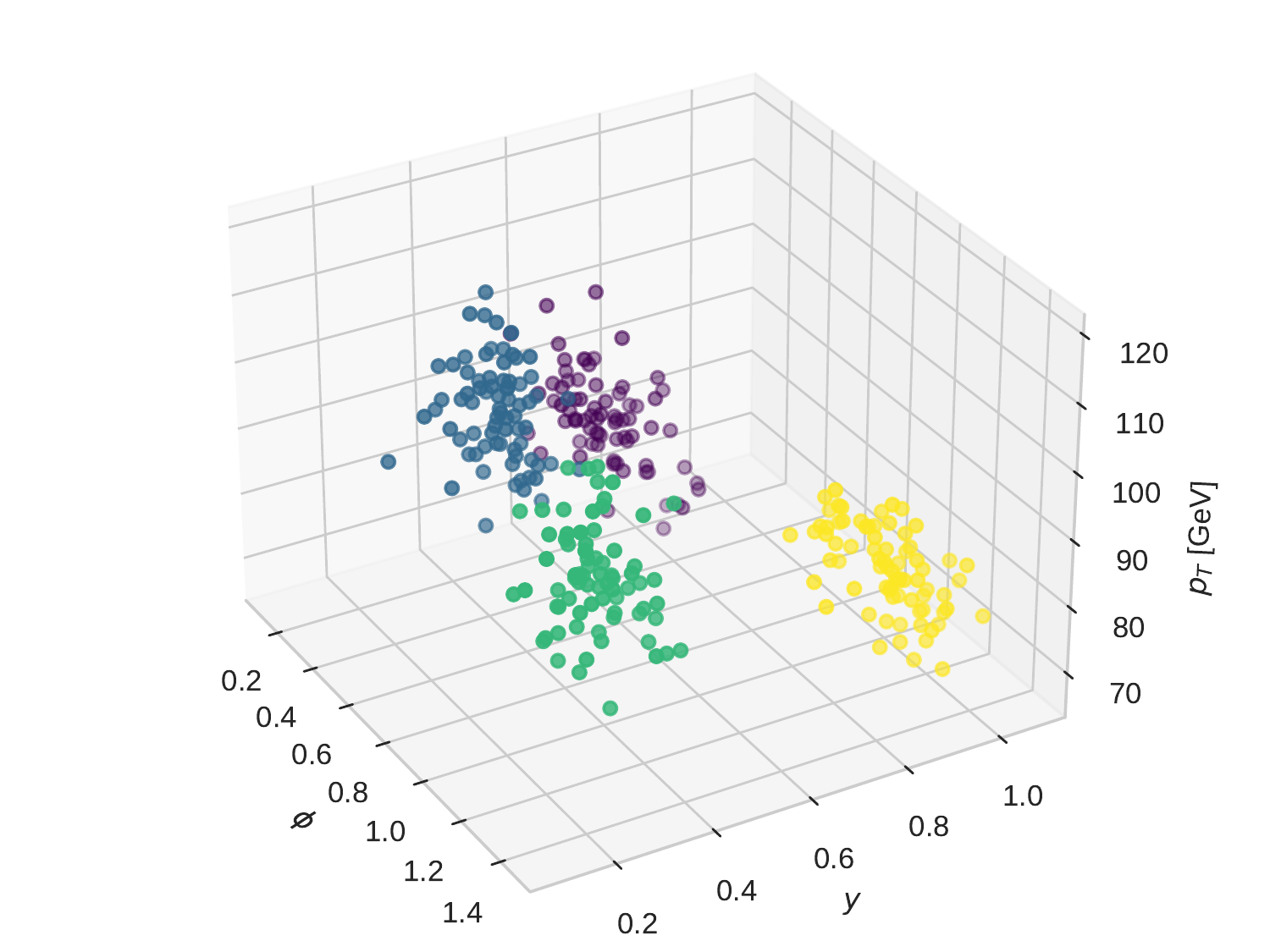}  
  \caption{ Classical \texttt{K-means} clustering, $\varepsilon_t=1.00$.}
  \label{fig:sub-first}
\end{subfigure}
\begin{subfigure}{.49\textwidth}
  \centering
  \includegraphics[width=.99\linewidth]{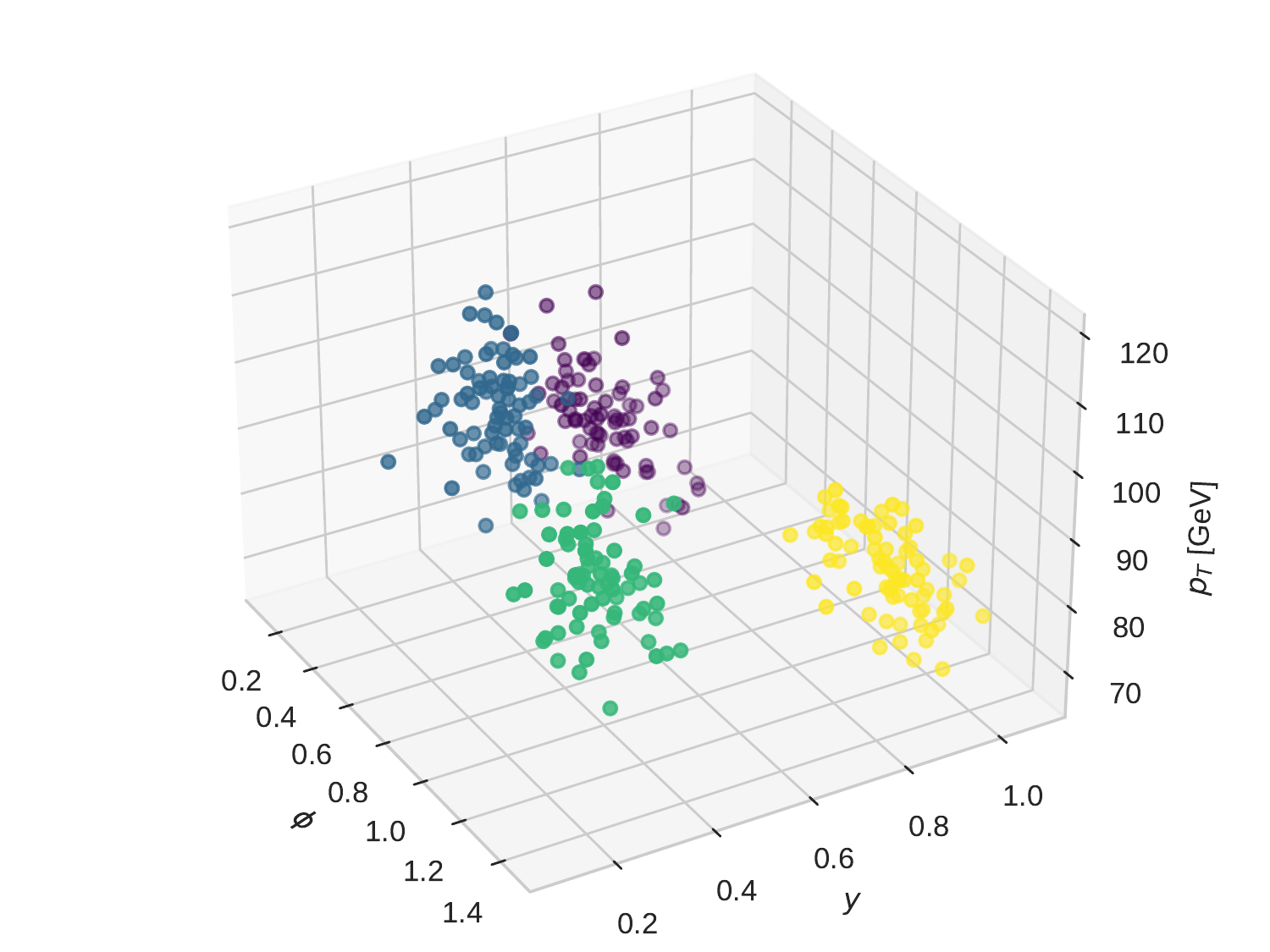}  
  \caption{ Quantum \texttt{K-means} clustering, $\varepsilon_t=1.00$.}
  \label{fig:sub-second}
\end{subfigure}
        \caption{In different colors, clusters identified after $5$ iterations by the classical and quantum versions of the \texttt{K-means} algorithm in a Gaussian dataset generated with a random seed and a standard deviation of $2.0$ from the cluster centroids.
        Note that clusterization has been performed using a Minkowski-type distance assuming that all particles are massless and on shell and the efficiencies of both algorithms are $\varepsilon_t = 1.00$. }
        \label{fig:qandckmeans}
\end{figure}

\begin{figure}[ht!]
       \centering
\begin{subfigure}{.49\textwidth}
  \centering
  \includegraphics[width=.99\linewidth]{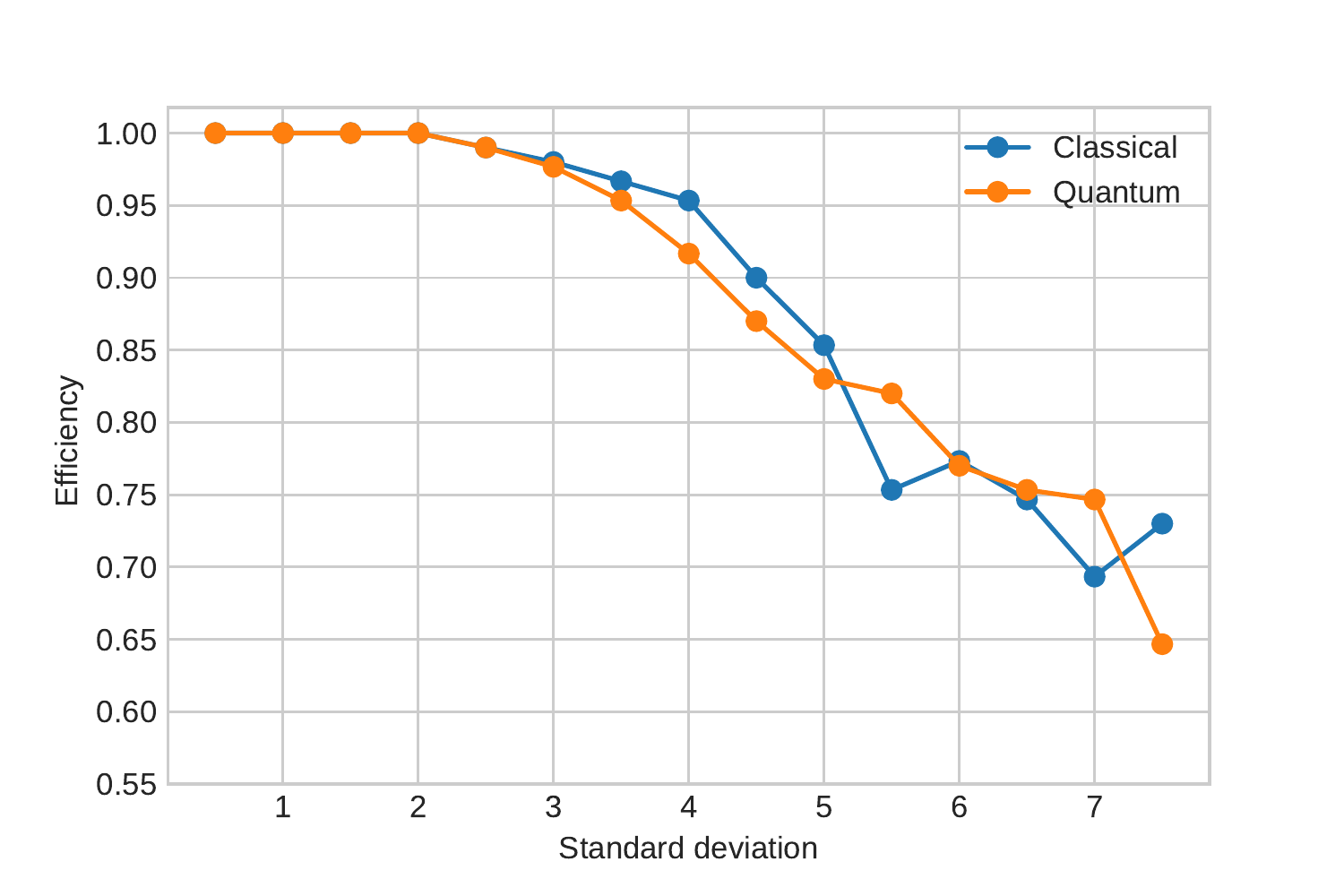}  
  \caption{Random seed.}
  \label{fig:efvsdeviation_sub-first}
\end{subfigure}
\begin{subfigure}{.49\textwidth}
  \centering
  \includegraphics[width=.99\linewidth]{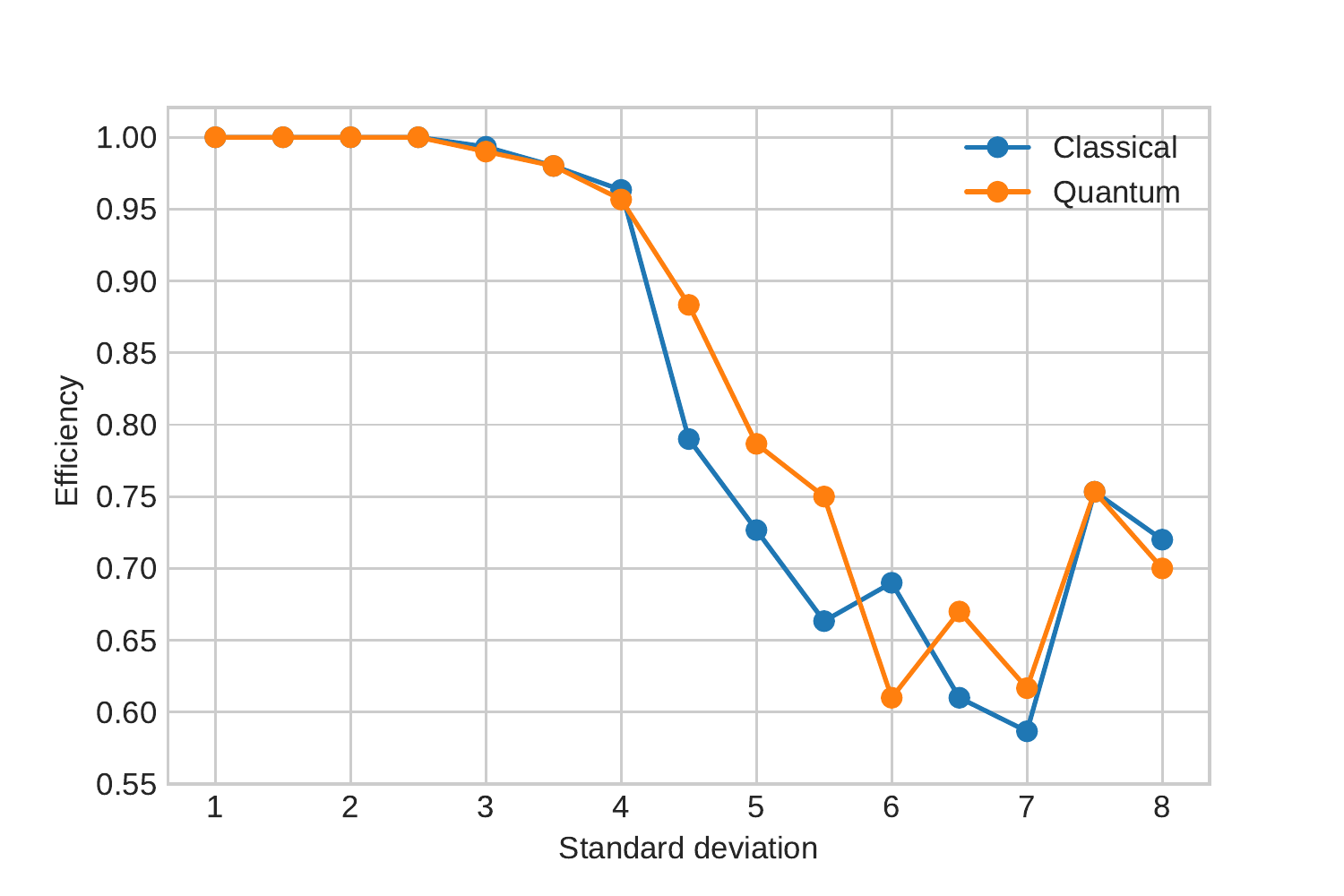}  
  \caption{\texttt{K-means}$++$ seed.}
  \label{fig:efvsdeviation_sub-second}
\end{subfigure}
        \caption{Cluster efficiency of the \texttt{K-means} algorithm versus standard deviations of the data with respect to centroids. Both the classical and quantum versions have been run on $15$ datasets with standard deviations ranging from $0.5$ to $7.5$. }
        \label{fig:efvsdeviation}
\end{figure}

At this point we present our implementation of the \texttt{K-means} algorithm with the invariant sum squared as a distance as well as a maximum searching algorithm, and compare its performance with its classical analogue.
To this end, we have generated 15 Gaussian clustered datasets of $N=300$ three-dimensional vectors~\footnote{In general, it is possible to relate this generated set of three-dimensional vectors, to a physical event at the LHC. It is enough to consider the set of $n$ three-dimensional vectors as massless partons recoiling against a small number of tagged particles.} with different levels of noise and clustering using the \textit{Scikit$-$learn} function \textit{make$\_$blobs}, which gives us the \textit{true labels} \footnote{The data generator function pre-assigns each data point to a particular cluster, so by analysing these \textit{true labels} one may know which is the correct way to cluster the data.} of the generated data. These \textit{true labels} of the data points are used to calculate the true efficiencies, $\varepsilon_t$, of the algorithms when analysing Gaussian datasets. The efficiency $\varepsilon_t$ is obtained as the ratio of the number of particles classified by the algorithm in the same way as the \textit{true labels} to the total number of particles. We then applied the hybrid and classical versions of the \texttt{K-means} algorithm to each dataset. Note that the data we are analysing represent the particle four-momenta in such a way that the three-dimensional vectors correspond to the spatial components, while the temporal components are calculated assuming that all particles are massless and on shell. 
Results are shown in Figs.~\ref{fig:qandckmeans} and~\ref{fig:efvsdeviation}.

\begin{figure}[ht!]
       \centering
\begin{subfigure}{.49\textwidth}
  \centering
  \includegraphics[width=.99\linewidth]{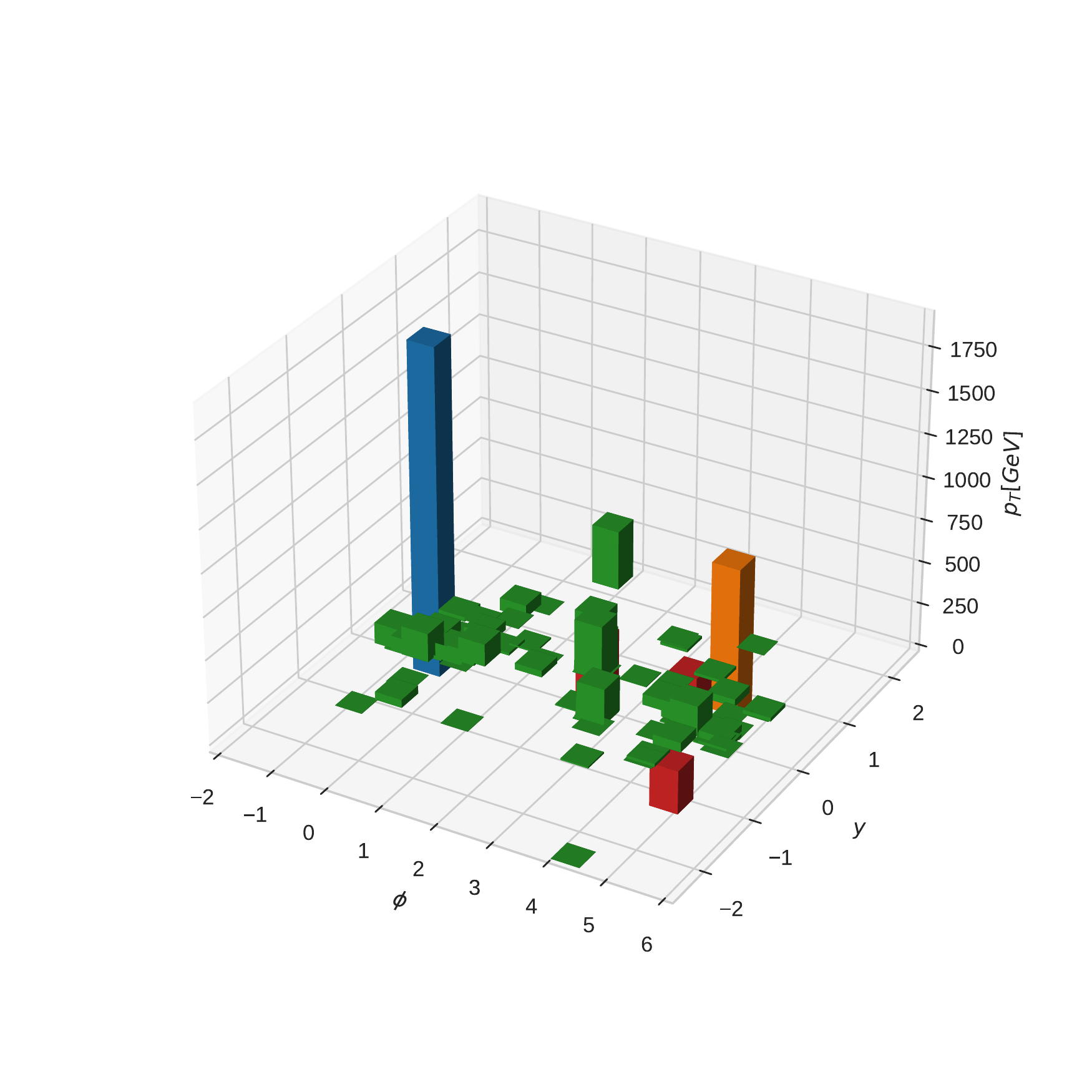}  
  \caption{\centering{Classical \texttt{K-means} applied to \hspace{\textwidth} LHC physical events.}}
  \label{fig:sub-first}
\end{subfigure}
\begin{subfigure}{.49\textwidth}
  \centering
  \includegraphics[width=.99\linewidth]{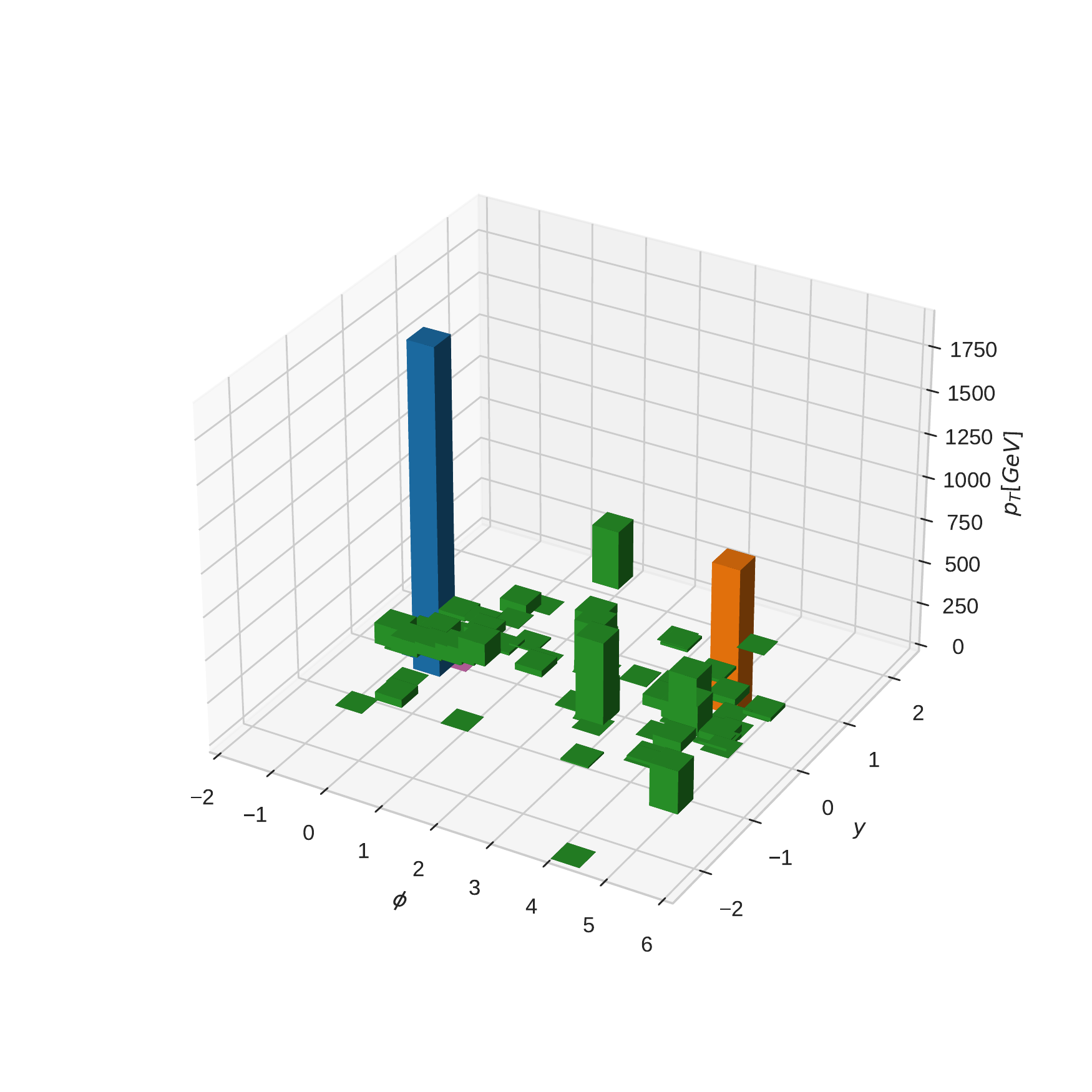}  
  \caption{\centering{Quantum \texttt{K-means} applied to \hspace{\textwidth} LHC physical events, $\varepsilon_c=0.94$.}}
  \label{fig:qkmeansrealdata_sub-second}
\end{subfigure}
        \caption{A sample parton-level event generated as described in the text and clustered with the classical and the quantum version of the \texttt{K-means++} algorithm, taking $K=8$.}
        \label{fig:qkmeansrealdata}
\end{figure}

Regarding Fig.~\ref{fig:qandckmeans} one can see at a glance that both classical and quantum versions perform the clustering in the same way in the three-dimensional space of transverse momentum ($p_T$), rapidity ($y$) and azimuth ($\phi$).

Fig.~\ref{fig:efvsdeviation} shows the efficiency in the reconstruction of the clusters as a function of the standard deviations used to generate the data, namely we check whether clustering occurs as expected. It is evident that for small values of the standard deviation both algorithms perform really well, with efficiencies close to one, while for larger values of the standard deviation (i.e. highly noisy data) both efficiencies drop. Furthermore, we can compare the performances of the \texttt{K-means} algorithm when the seed of the centroids is chosen randomly (see Fig. \ref{fig:efvsdeviation_sub-first}), with respect to the case when the seed centroids are carefully selected to be as far as possible from each other, according to the \texttt{K-means}$++$ prescription (see Fig. \ref{fig:efvsdeviation_sub-second}). The random seed variant in Fig. \ref{fig:efvsdeviation_sub-first}, has a linear decrease with respect to the standard deviation, and the performances of classical and quantum versions are very similar. On the other hand, the \texttt{K-means}$++$ variant, Fig. \ref{fig:efvsdeviation_sub-second}, presents a different behaviour. The quantum version outperforms, in the majority of the cases, the classical one from a standard deviation of 4 onward. Furthermore, in this variant both performances show a dropoff from 4 standard deviations to 7, and then a slight rise from 7 to 8. Finally, comparing both variants it is observed that the \texttt{K-means}$++$ method outperforms the random seed case for small values of the standard deviation ($<4$). However, for larger values of the standard deviation the random seed prescription presents higher efficiencies.

In the following, we will apply our quantum \texttt{K-means} method to LHC physical events.
To do so we first have processed the data to avoid the following problem: a negative vector $-{\bf x}$ represents the same quantum state $|x\rangle$ as its positive analogue ${\bf x}$ up to a global phase. 
This data processing consists of rescaling the data to be analysed in the interval \{1,10\}\footnote{Note that the value 0 is not included to avoid numerical and statistical fluctuations}. This means every component of every data point will be rescaled in the desired interval. Thus, all the data points are positive now. Moreover, when analysing LHC physical events, we no longer have the \textit{true labels}, so we cannot calculate $\varepsilon_t$. Instead, we define the efficiency $\varepsilon_c$, which is defined as the quotient of the number of particles clustered in the same way as their classical counterpart and the total number of particles to be classified.

We consider the generation of a physical $n$-particle event produced at the LHC. We use a private implementation of an $n$-particle ($n$ can be of the order of tens of thousands) phase-space event generator. This \texttt{C++} code, which is based on \texttt{ROOT} \cite{Brun:1997pa}, generates $n$-particle events, in which the final-state particles can be massive or massless in any combination of each other (combination chosen by the user). This allows the user to generate final states in which all the particles are massless QCD partons, massless QCD partons associated with photons, massive vector bosons, top-quarks, etc.

The precision in the generation of the final-state event is verified on an event-by-event basis by computing the kinematical constraint between the initial and the $n$-particle final state. The required precision~\footnote{If we consider all momenta of the event outgoing, the kinematical constraint is evaluated over the resulting three-momentum space vector. The test in the accuracy of the kinematical constraint is performed at the highest multiplicity in the final state. This constitutes the lowest limit for the precision, since reducing the particle number in the final state, the precision improves.} is always better than $10^{-2}$. 
Each generated event is then analysed with the classical versions of the $k_T$-jet algorithms (as implemented in \texttt{FastJet} \cite{Cacciari:2011ma}) and with our quantum version of the corresponding jet algorithms.

In this paper we consider the $n$-particle massless final-state production in proton-proton~\footnote{Since we are considering unweighted events, our study is not only valid for proton-proton colliders, but also for $e^{+}e^{-}$ colliders.} collisions at a centre-of-mass energy of $\sqrt{s}$ = 14 TeV. We apply the following final-state selection cuts. We select jets with the $k_T$-jet algorithms according to the following parameters: the minimum transverse momentum of the resulting jets is required to be $p_{T {\rm min}} \geq $ 10 GeV and with a radius $R = 1$. For our study, we consider $n$ massless particles in the final state with $n=128$.

The application of the quantum \texttt{K-means++} method to LHC physical events is displayed in Fig. \ref{fig:qkmeansrealdata}.
Notice that even if we choose $K=8$ beforehand, one may see in Fig. \ref{fig:qkmeansrealdata} that the algorithms clearly distinguish only 3 or 4 clusters (jets). There is actually a simple explanation. Although the algorithm starts with $K$ centroids, the algorithm may converge to a local minimum when the number of clusters is less than $K$, leaving the remaining clusters completely empty.

In Fig.~\ref{fig:qkmeansrealdata} one can observe graphically that both algorithms classify the data in much the same way, and also the efficiency shown by the quantum algorithm is close to one. Therefore,  the results of this quantum version using physical data may be considered satisfactory.

\subsection{Quantum Affinity Propagation algorithm}
\label{subsec:qapalgorithm}

In this subsection, a simulation of the quantum \texttt{AP} algorithm is presented. First, we apply this algorithm to Gaussian datasets with different numbers of clusters, generated with a standard deviation of 0.6. That value of the standard deviation has been chosen arbitrarily by convenience. The efficiencies resulted for the classical and the quantum versions are shown in Table \ref{tab:qapclusters}.
Table~\ref{tab:qapclusters} depicts that the \texttt{AP} classical algorithm and its quantum counterpart clustered the low-noise Gaussian datasets successfully.

\begin{table}[th!]
\begin{longtable}{| p{3cm} | p{3cm} | p{3cm}|}
\hline
\centering {Number of clusters $K$}& \centering{ Efficiency classical \texttt{AP}} ($\varepsilon_t)$ & \centering {Efficiency quantum \texttt{AP} ($\varepsilon_t)$} \cr   \hline 
 \centering 4 &\centering 1.00 &\centering 0.99 \cr   \hline
\centering 5 &\centering 1.00 &\centering 1.00 \cr   \hline
\centering 6 &\centering 0.99 &\centering 0.98 \cr   \hline
\centering 7 &\centering 1.00 &\centering 0.98  \cr   \hline
\centering 8 &\centering 0.98 &\centering 0.94  \cr    \hline 
\omit
    \\    
\caption{Efficiencies of \texttt{AP} algorithms for Gaussian datasets with different number of clusters.}
\label{tab:qapclusters}
\end{longtable}
\end{table}

\begin{figure}[ht!]
       \centering
\begin{subfigure}{.49\textwidth}
  \centering
  \includegraphics[width=.99\linewidth]{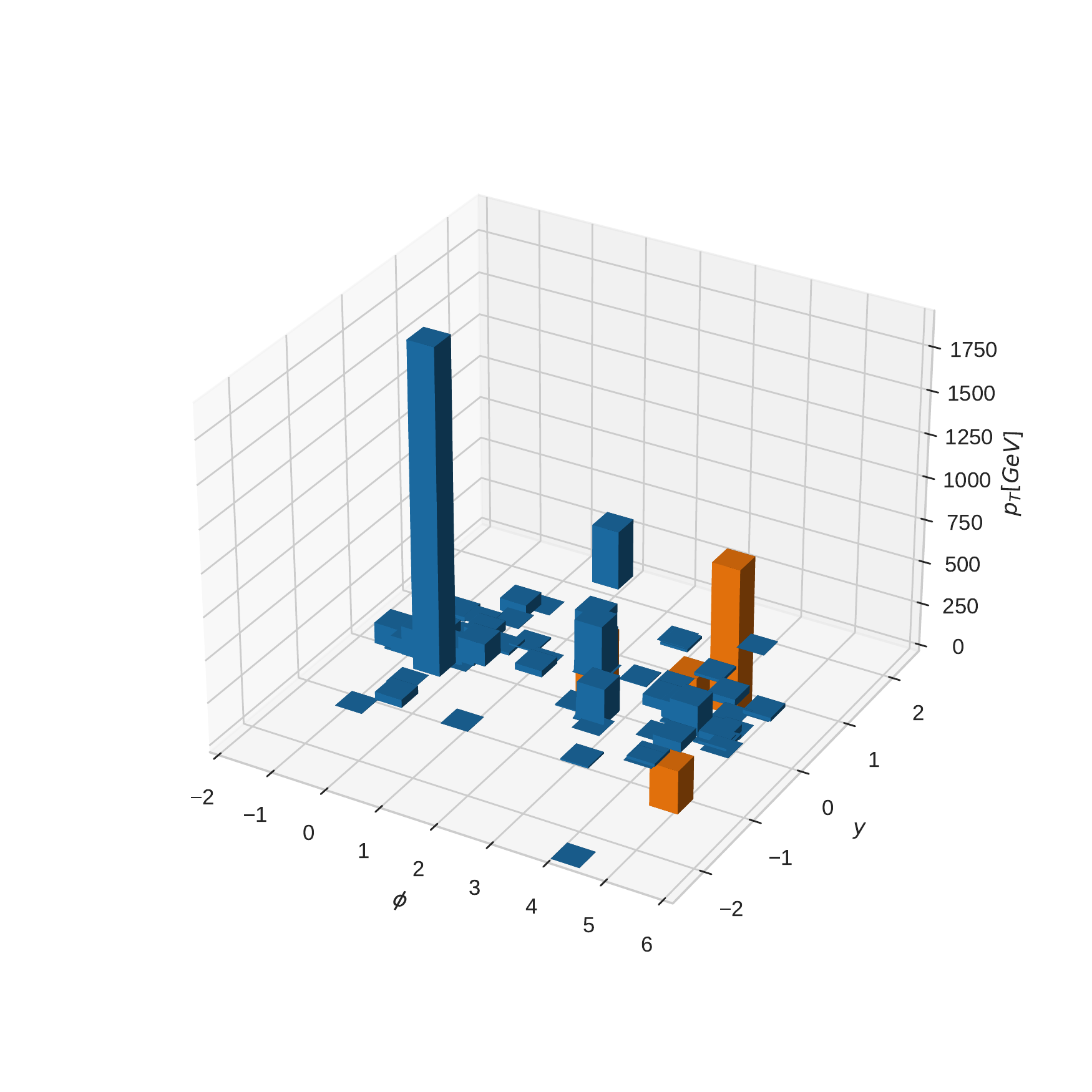}  
  \caption{\centering Classical \texttt{AP} algorithm applied to \hspace{\textwidth} LHC physical events.}
  \label{fig:applots_sub-first}
\end{subfigure}
\begin{subfigure}{.49\textwidth}
  \centering
  \includegraphics[width=.99\linewidth]{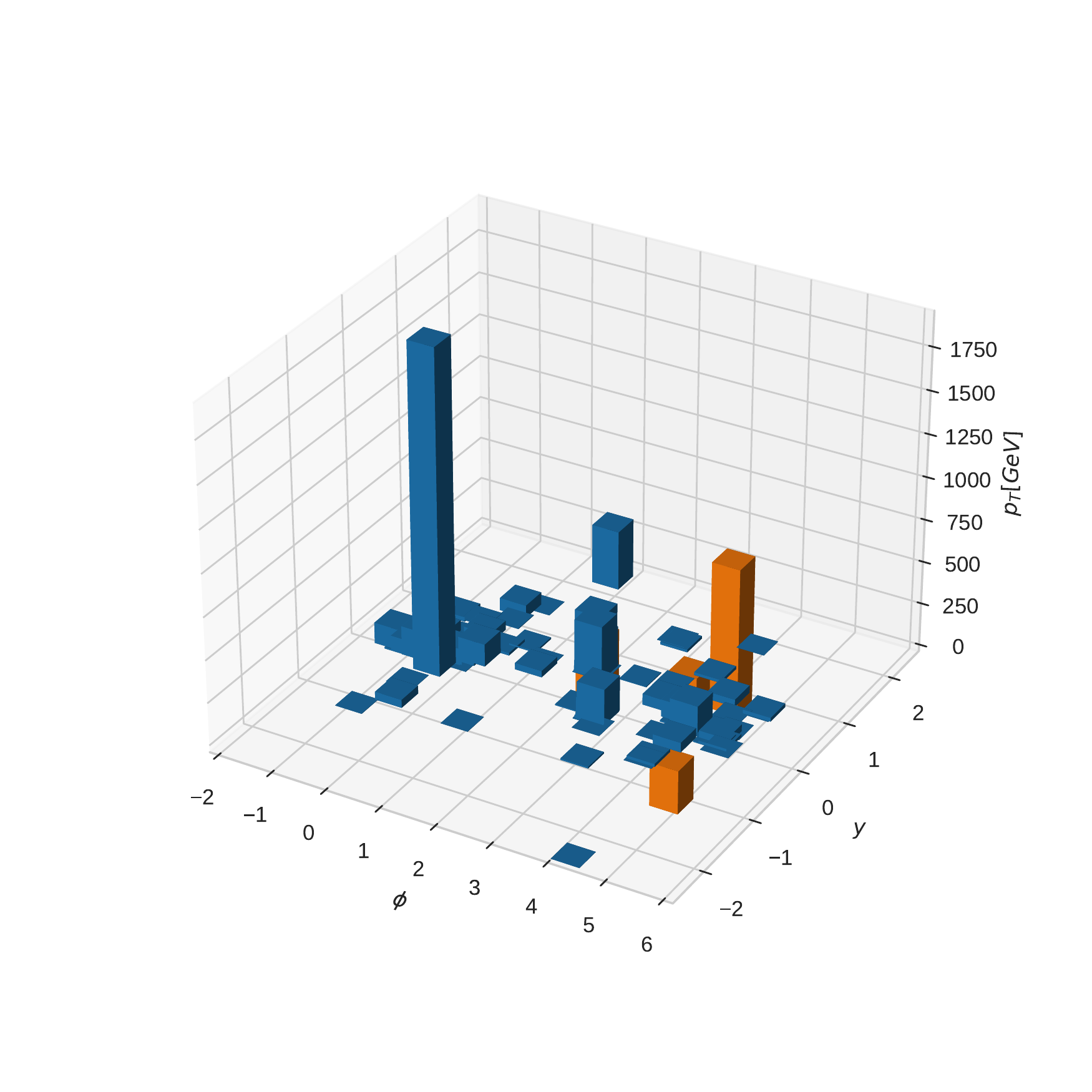}  \caption{\centering Quantum \texttt{AP} algorithm applied to \hspace{\textwidth} LHC physical events, $\varepsilon_c=1.00$.}
  \label{fig:applots_sub-second}
\end{subfigure}
        \caption{A sample parton-level event generated as described in the text and clustered in $K=2$ different clusters with the classical and the quantum version of the \texttt{AP} algorithm. }
        \label{fig:applots}
\end{figure}

In the following, we apply this algorithm to the physical dataset described in Section~\ref{subsec:qkmeans}, which was preprocessed  for the reasons explained in the same section. The results obtained are shown in Fig. \ref{fig:applots}.
In Fig. \ref{fig:applots_sub-second} exactly the same clustering is performed as in Fig. \ref{fig:applots_sub-first} (notice that the efficiency of the quantum version is $\varepsilon_c=1.00$). 
Nonetheless, this algorithm only finds 2 clusters, which differs with respect to the 3 or 4 clusters found by the \texttt{K-means} algorithm (see Fig. \ref{fig:qkmeansrealdata}). Even more, both algorithms identify correctly the most energetic jets of the event (the blue and the orange ones) while the majority of the remaining particles are not classified in the same way, probably because they are soft particles.

\subsection{Quantum $k_T$ jet algorithm}
\label{subsec:qktalgorithm}

In this section, we apply the quantum version of the $k_T$-jet algorithm to the same LHC physical events as described in Section~\ref{subsec:qkmeans} in order to compare the three clustering algorithms. 

\begin{figure}[H]
\begin{subfigure}{.5\textwidth}
  \centering
  \includegraphics[width=.8\linewidth]{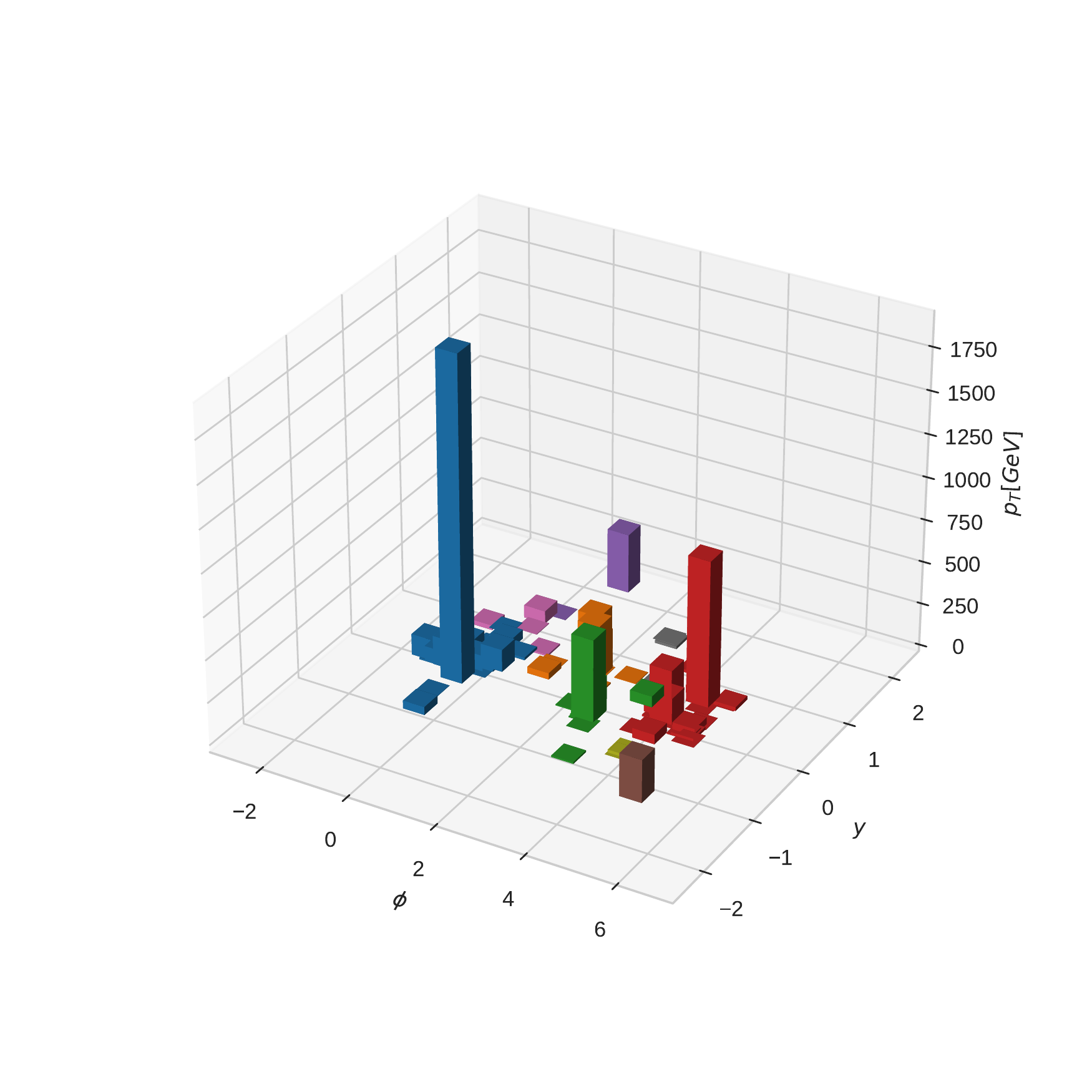}  
  \caption{Classical anti-$k_T$, $p=-1$, $R=1$.}
  \label{fig:sub-first}
\end{subfigure}
\begin{subfigure}{.5\textwidth}
  \centering
  \includegraphics[width=.8\linewidth]{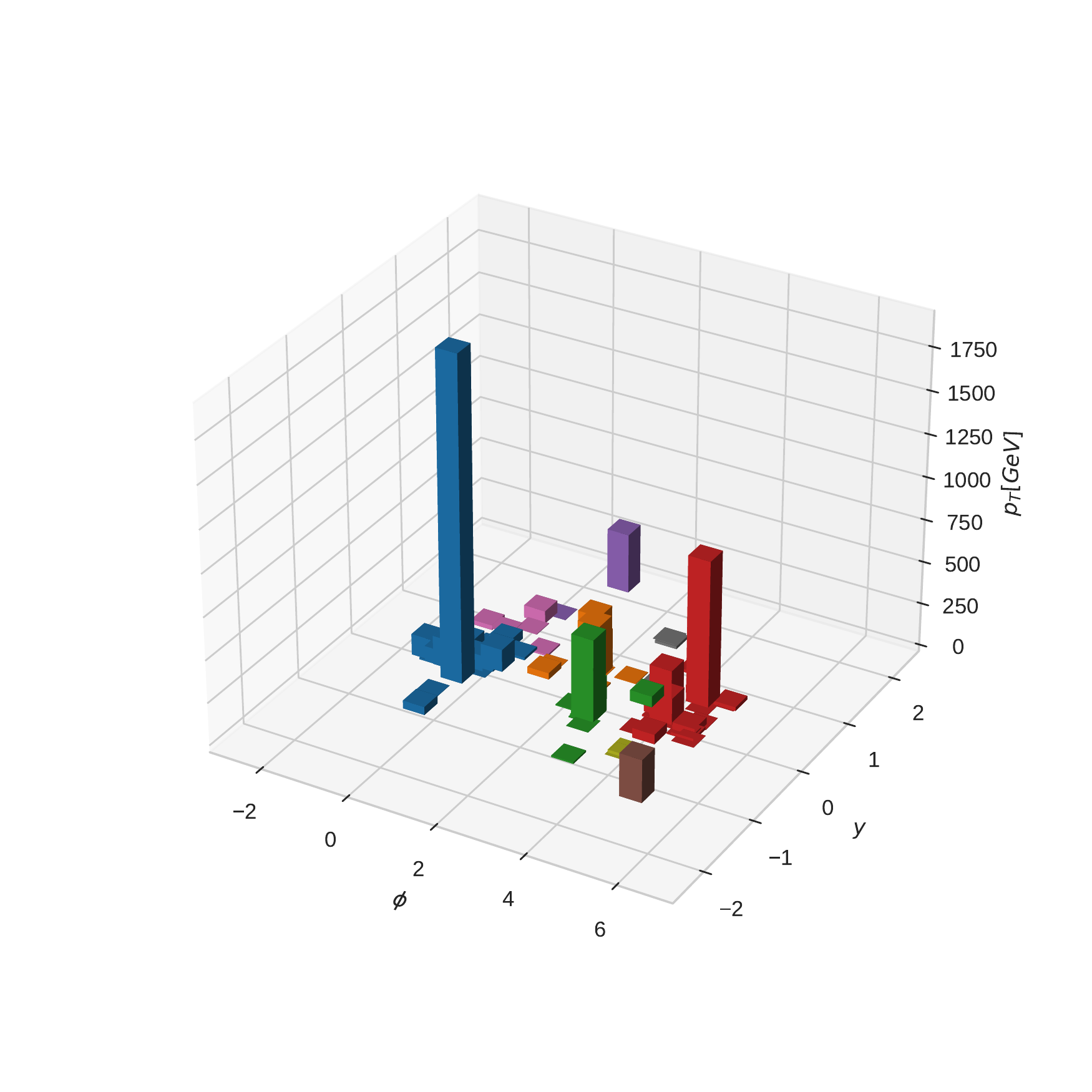}  
  \caption{Quantum anti-$k_T$, $p=-1$, $R=1$, $\epsilon_c =0.99$.}
  \label{fig:sub-second}
\end{subfigure}
\newline
\begin{subfigure}{.5\textwidth}
  \centering
  \includegraphics[width=.8\linewidth]{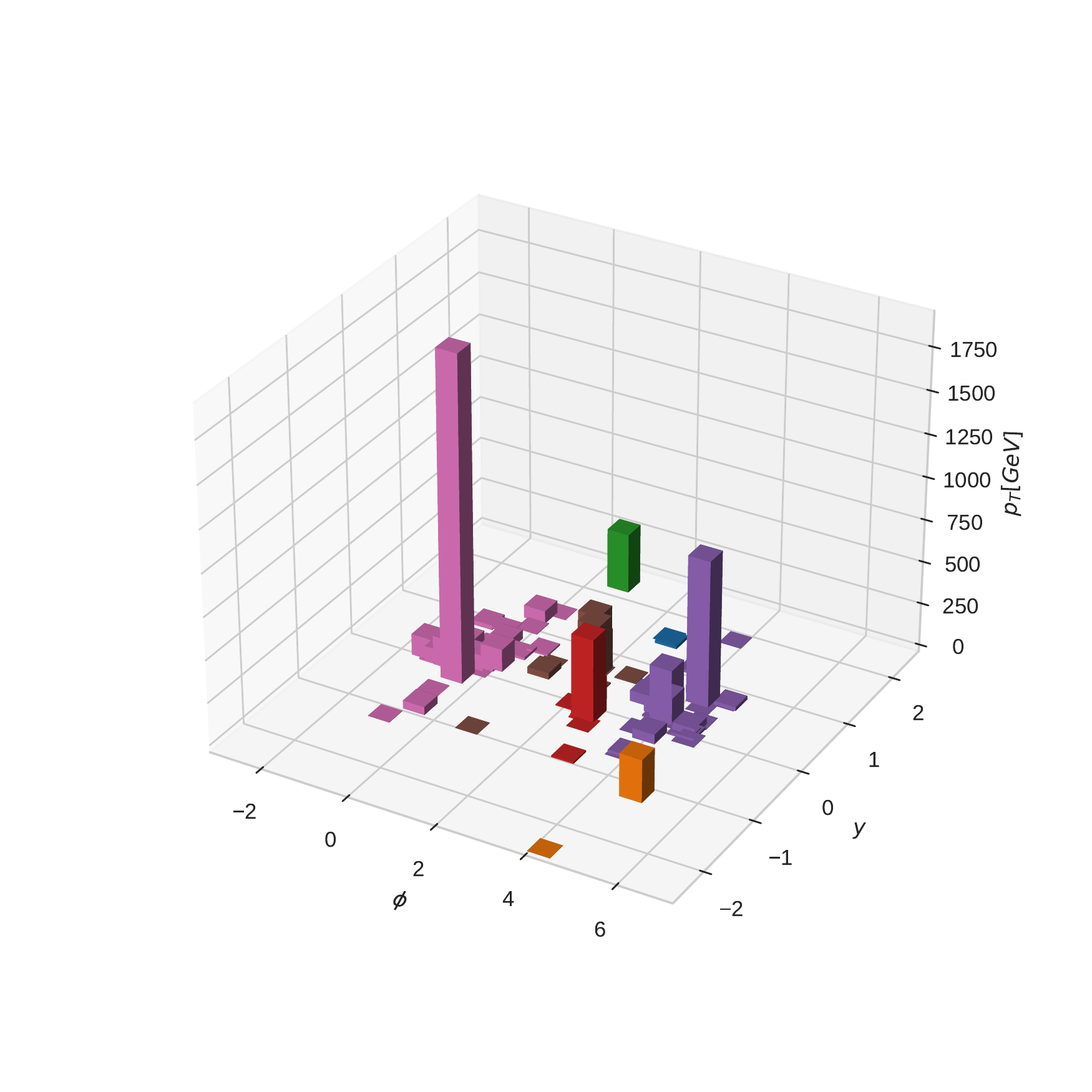}  
  \caption{Classical $k_T$, $p=1$, $R=1$.}
  \label{fig:sub-third}
\end{subfigure}
\begin{subfigure}{.5\textwidth}
  \centering
  \includegraphics[width=.8\linewidth]{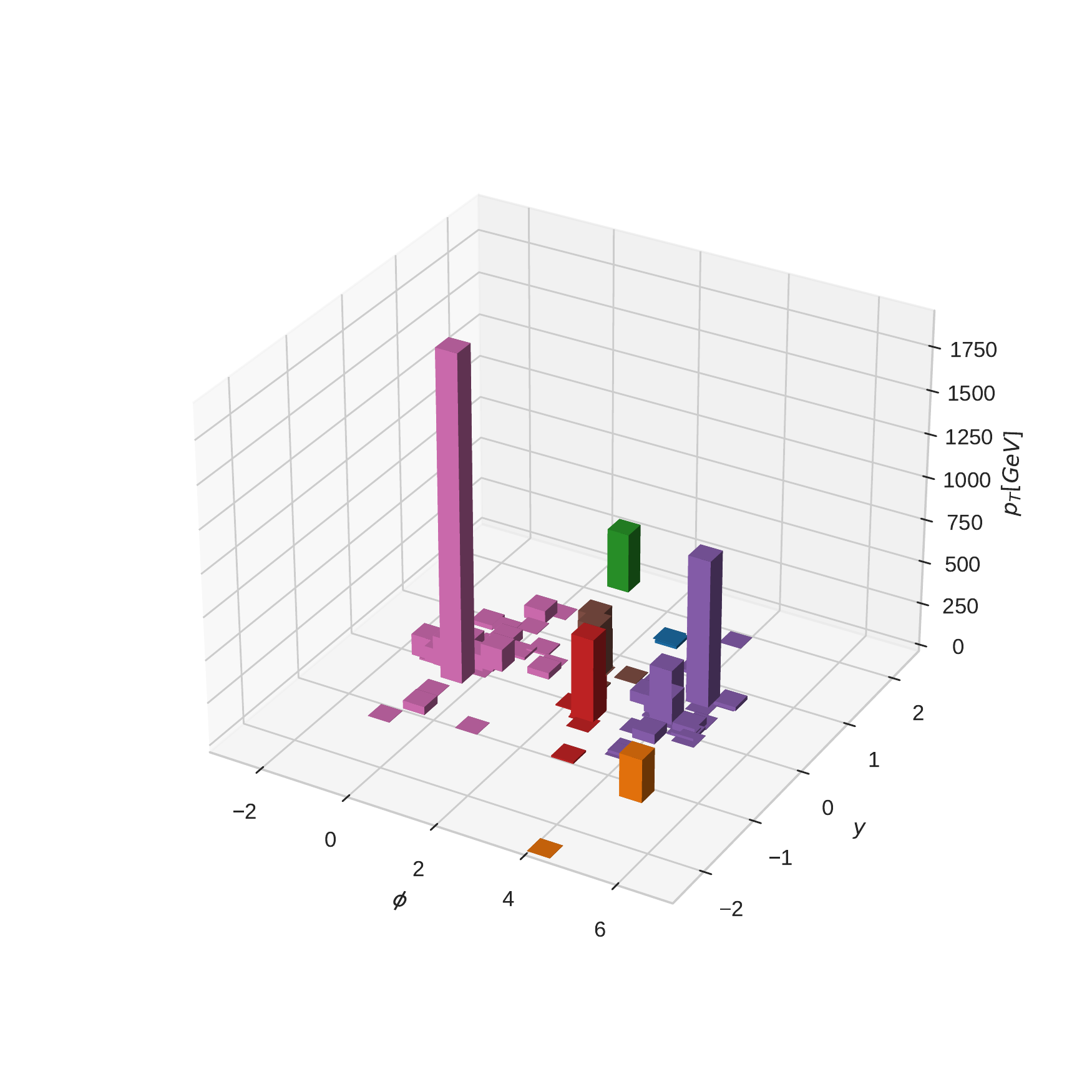}  
  \caption{Quantum $k_T$, $p=1$, $R=1$, $\epsilon_c =0.98$.}
  \label{fig:sub-fourth}
\end{subfigure}
\newline
\begin{subfigure}{.5\textwidth}
  \centering
  \includegraphics[width=.8\linewidth]{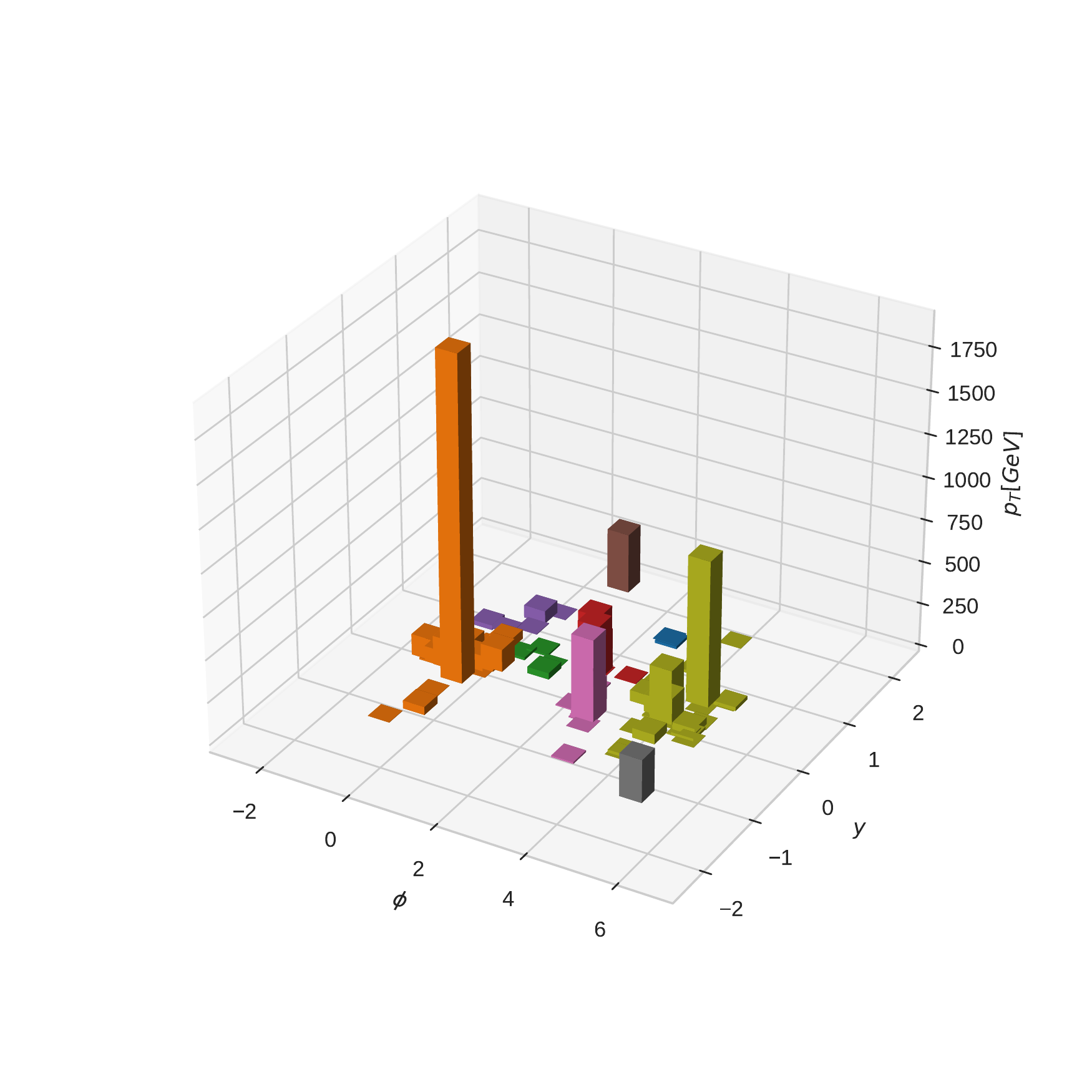}  
  \caption{Classical Cam/Aachen, $p=0$, $R=1$.}
  \label{fig:sub-fifth}
\end{subfigure}
\begin{subfigure}{.5\textwidth}
  \centering
  \includegraphics[width=.8\linewidth]{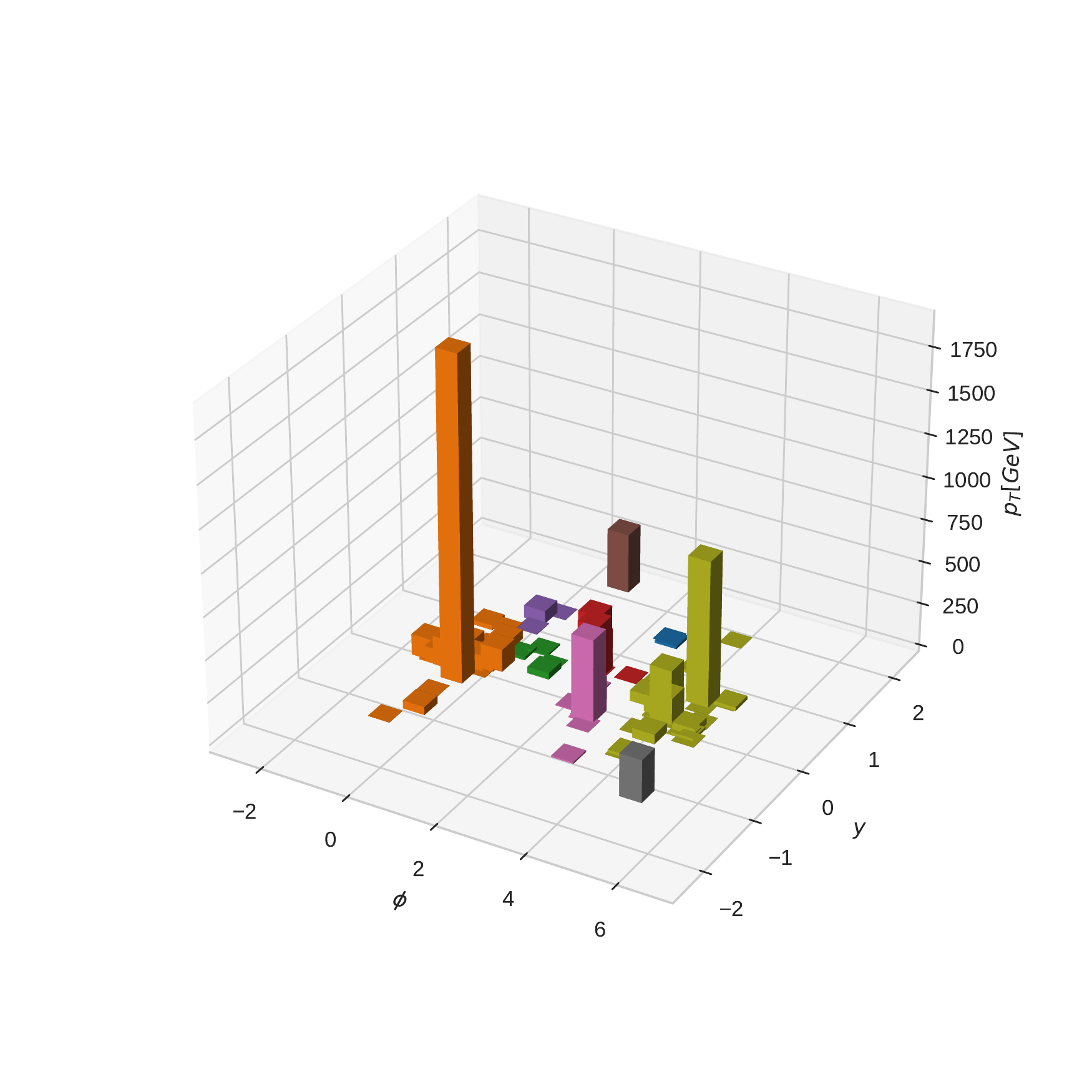}  
  \caption{Quantum Cam/Aachen, $p=0$, $R=1$, $\epsilon_c =0.98$.}
  \label{fig:sub-sixth}
\end{subfigure}
\caption{ A sample parton-level event generated as described in the text and clustered with three different $k_T$-jets algorithms as well as its quantum versions.} 
\label{fig:ktclassicandquantum}
\end{figure}

In Fig.~\ref{fig:ktclassicandquantum} we show the performance of classical and quantum $k_T$ jet algorithms. It depicts the jet clustering process carried out by each one of the $k_T$ algorithm versions, i.e. anti-$k_T$, $k_T$ and Cambridge/Aachen. The classical and quantum versions perform the same jet clustering.
When comparing Figs.~\ref{fig:qkmeansrealdata}, \ref{fig:applots} and \ref{fig:ktclassicandquantum}, one can observe that the latter performs a cleaner clusterization with a larger number of jets.
This is a visual effect because jet clusterization is represented graphically in 3-dimensions, which coincides with the dimensionality of the $k_T$ metrics, while the  \texttt{K-means} and \texttt{AP} use a 4-dimensional Minkowski distance.


To conclude this section we also analyse the efficiencies and the number of \textit{shots} required for all the quantum versions as a function of the $a$ parameter (see Section \ref{subsec:ktjetalgorithm}). These are shown in Table~\ref{tab:effsandshots}. 
Table~\ref{tab:effsandshots} displays that the efficiencies of the quantum algorithms are close to one, i.e., they classify particles almost identically to their classical counterparts. 
Furthermore, it may be observed that the larger the parameter $a$, the smaller the number of \textit{shots} required to achieve a successful efficiency. 
In this case, we only need to increase the parameter $a$ to the number $5$ to achieve the desired efficiencies with at most $10$ \textit{shots}. However, in other problems (with a larger dataset) a parameter greater than $a=5$ can be used to separate the data points and achieve the highest possible efficiency with the smallest number of \textit{shots}.

\begin{table}[th!]
\begin{longtable}{| p{0.5cm} | p{1.9cm} | p{1.8cm} | p{2.3cm} | p{1.7cm} | p{2.3cm} | p{2.3cm} |}
\hline
\centering {$a$}& \centering{ Efficiency anti-$k_T$} & \centering {\textit{Shots} anti-$k_T$} & \centering {Efficiency $k_T$} & \centering{\textit{Shots} $k_T$} & \centering {Efficiency Cam/Aachen} &  \centering{  \textit{Shots} Cam/Aachen} \cr   \hline 
 \centering 1 &\centering 0.96 &\centering 50 &\centering 0.98 &\centering 50 &\centering 0.96 &\centering 70\cr   \hline
\centering 2 &\centering 0.99 &\centering 40 &\centering 0.99 &\centering 45 &\centering 0.98 &\centering 60 \cr   \hline
\centering 3 &\centering 1.00 &\centering 25 &\centering 0.98 &\centering 20 &\centering 0.97 &\centering 40\cr   \hline
\centering 4 &\centering 1.00 &\centering 15 &\centering 0.95 &\centering 15 &\centering 1.00 &\centering 20 \cr   \hline
\centering 5 &\centering 0.99 &\centering 5 &\centering 1.00 &\centering 8 &\centering 0.98 & \centering 10 \cr    \hline 
\omit
    \\    
\caption{Efficiencies and number of \textit{shots} of the different quantum $k_T$-jet algorithms as a function of parameter $a$.}
\label{tab:effsandshots}
\end{longtable}
\end{table}

\section{Conclusions}
\label{sec:conclu}

In this paper, we have considered the quantum versions of the well-known \texttt{K-means}, Affinity Propagation and $k_T$-jet clustering algorithms.
These quantum versions are based on two novel quantum procedures. The first one is a quantum subroutine which serves to compute distances satisfying Minkowski metric, whereas the second one consists of a quantum circuit to track the maximum into a list of unsorted data.

In the case of the \texttt{K-means} clustering algorithm, the quantum version is based on the standard classical algorithm with a quantum procedure to compute distances in Minkowski space and an additional quantum procedure to assign each particle to the nearest centroid. We found that the \texttt{K-means} quantum algorithm has a clustering efficiency as good as its classical counterpart while it would show an exponential speed-up in computational time in the vector dimensionality $d$, as well as in the number of clusters $K$ on a quantum device with qRAM.

In the second place, we have considered a quantum version of the Affinity Propagation method, which is an unsupervised machine learning algorithm, where the similarity is computed with the same quantum procedure as in the \texttt{K-means} case. Thus, it would lead to an exponential speed-up regarding its classical counterpart in the vector dimensionality $d$ while maintaining the clustering efficiency.

Finally, we have presented the quantum versions of the well-known $k_T$-jet clustering algorithms. On a true universal quantum device, the implementation of these algorithms would exhibit an exponential speed-up in finding the minimum distance.
Therefore, while the classical version requires $\mathcal{O}(N^3)$ in computational cost, where $N$ is the number of particles to cluster, the quantum counterpart would only require $\mathcal{O}(N^2\log(N))$. Notice that this comparison is performed between the classical \textit{non-optimal} and not optimized version and its quantum analogue. Further improvements can be obtained by applying to the quantum algorithm the geometrical nearest neighbour optimization procedure that is also applied to \texttt{FastJet}. In this way, we would obtain a quantum version of order $\mathcal{O}(N\log(N))$, which is of the same order as the fully optimal version of \texttt{FastJet}.

For all the clustering algorithms considered, the quantum simulations presented in this paper show an excellent performance and clustering efficiencies. Furthermore, the comparison with their classical counterparts displays that both classifications of the LHC simulated data are quite in agreement.


\section*{Acknowledgements}
We thank Abhijat Sharma and Guillermo Alonso for very helpful conversations and suggestions.
This work is supported by the Spanish Government (Agencia Estatal de Investigaci\'on MCIN/AEI/ 10.13039/501100011033) Grant
No. PID2020-114473GB-I00, and Generalitat Valenciana
Grant No. PROMETEO/2021/071. LC is supported by Generalitat Valenciana
GenT Excellence Pro\-gramme (CIDEGENT/2020/011).


\bibliography{bibliography}

\appendix
\setcounter{section}{0}
\section{Controlled \textit{SwapTest}}
\label{app:swaptest}
\begin{figure}[ht!]
    \centering
    \includegraphics[width=0.45\textwidth]{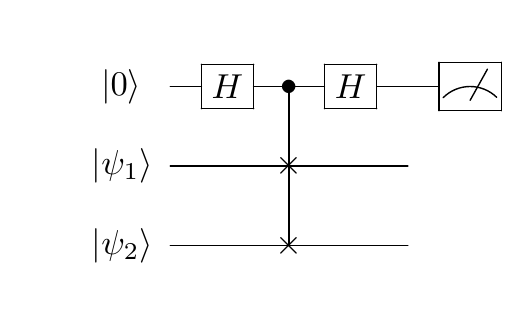}
    \caption{Quantum circuit \textit{SwapTest}.}  
    \label{fig:swaptest}
\end{figure}

A well-known procedure for determining the entanglement between two quantum states is the controlled \textit{SwapTest}~\cite{Buhrman:2001} method. This method allows us to quantify the overlap between $\ket{\psi_1}$ and $\ket{\psi_2}$, which are two input general quantum states of $n$ and $m$ qubits respectively such that $n\geq m$ (otherwise we exchange the labels $1$ and $2$), by measuring an ancillary qubit.
The controlled \textit{SwapTest} proceeds in three steeps starting from the initial state
\beq
\ket{\Psi_0} = \ket{0, \psi_1, \psi_2}~,
\eeq
where the ancillary qubit has been initialized to $\ket{0}$. 
In the first step, a Hadamard ($H$) gate is applied to the ancillary qubit,
while the states to be probed are left unchanged, resulting 
in the new state
\beq
\ket{\Psi_1} = \left( H \otimes \id^{\otimes n+m}\right) \ket{\Psi_0}
=\frac{1}{\sqrt{2}} \left( \ket{0 ,\psi_1, \psi_2}
+\ket{1 ,\psi_1, \psi_2}\right)\,,
\eeq
where the identity $\id^{\otimes n+m}$ acts over the $\ket{\psi_1}$ and $\ket{\psi_2}$ states and the tensor product $\otimes$ is omitted in the composed states (e.g $\ket{0} \otimes \ket{\psi_1} \otimes \ket{ \psi_2} = \ket{0 ,\psi_1, \psi_2}$).
A controlled swap gate (CSWAP) is then applied to $\ket{\Psi_1}$ where all the $m$ qubits of $\ket{\psi_2}$ are swapped with the $m$ first qubits of $\ket{\psi_1}$, leading to 
\beq
\ket{\Psi_2} = {\rm CSWAP} \ket{\Psi_1}
=\frac{1}{\sqrt{2}} \left( \ket{0, \psi_1, \psi_2}
+\ket{1 ,\psi_2 ,\psi_1'}\right)~,
\eeq
where $\psi_i'$, is the swapped state of $\psi_i$, i.e., a state where the $m$ first qubits of $\psi_1$ have been swapped with the rest $n-m$ qubits. The final step consist of 
applying again a Hadamard gate to the ancillary qubit
\beq
\ket{\Psi_3} = \left( H \otimes \id^{\otimes n+m}\right) \ket{\Psi_2}
=\frac{1}{2} \left( \ket{0} \otimes \left(\ket{\psi_1 ,\psi_2} + \ket{\psi_2, \psi_1'}\right)
+\ket{1} \otimes \left( \ket{\psi_1, \psi_2} - \ket{\psi_2, \psi_1'} \right)
\right)~.
\eeq
The resulting probability of measuring the ancillary qubit in the 
state $\ket{0}$ is given by 
\beq
\begin{split}
P_{\Psi_3}(\ket{0}) &= \left|\langle 0\ket{\Psi_3} \right|^2 = 
\frac{1}{4} \left| \ket{\psi_1, \psi_2} + \ket{\psi_2, \psi_1'} \right|^2 
=\frac{1}{2}+\frac{1}{2} {\rm Re} 
\left[ \la \psi_2, \psi_1' \ket{\psi_1, \psi_2} \right]\\
&=\frac{1}{2}+\frac{1}{2} \la \psi_1'| \psi_2 \ra \la \psi_2| \psi_1 \ra ~,
\end{split}
\label{eq:p0swaptest}
\eeq
which turns out to be as follows if $m=n$, thus $|\psi_1' \ra$=$|\psi_1 \ra$
\beq
P_{\Psi_3}(\ket{0}) =
\frac{1}{2}+\frac{1}{2} \left|\la \psi_1| \psi_2 \ra \right|^2~,
\label{eq:p0swaptestsimple}
\eeq
and this provides the squared inner product between 
the two states with an uncertainty of ${\cal O}(\epsilon)$
after ${\cal O}(\epsilon^{-2})$ shots. The corresponding quantum circuit associated to the \textit{SwapTest} method is shown in Fig.~\ref{fig:swaptest}.



\end{document}